\documentclass[11pt,reqno,final]{amsart}
\pdfoutput=1
\usepackage[round,sort,elide]{natbib}
\usepackage{graphicx}
\usepackage{times}
\usepackage{rotating}
\usepackage{subfig}
\usepackage{color}

\theoremstyle{plain}

\numberwithin{equation}{part}

\renewcommand\thesection{\arabic{section}}

\renewcommand\thefigure{\arabic{figure}}
\renewcommand\thetable{\arabic{table}}

\newcommand\scinot[2]{$#1 \times 10^{#2}$}

\newcommand{\dlta}[1]{{\Delta}{#1}}

\newcommand{\Expect}[1]{\mathbb{E}\left[#1\right]}

\newcommand{\dd}[1]{\mathrm{d}{#1}}

\IfFileExists{upquote.sty}{\usepackage{upquote}}{}
\begin{document}

\begin{flushright}
Version dated: \today
\end{flushright}
\bigskip
\noindent
Detecting adaptation using O-U models
\bigskip
\medskip
\begin{center}

\noindent
{\Large \bf Detecting adaptive evolution in phylogenetic comparative analysis using the Ornstein-Uhlenbeck model}
\bigskip

\noindent
{\normalsize \sc Clayton E. Cressler$^1$, Marguerite A. Butler$^2$, and Aaron A. King$^3$}\\
\noindent
{\small \it
$^1$Department of Biology, Queen's University, Kingston, ON, K7L 3N6, Canada;\\
$^2$Department of Zoology, University of Hawai'i, Honolulu, HI, 96822, USA;\\
$^3$Departments of Ecology \& Evolutionary Biology and Mathematics, University of Michigan, Ann Arbor, MI, 48109, USA}\\
\end{center}
\medskip
\noindent
{\bf Corresponding author:} Clayton E. Cressler, Department of Biology, Queen's University, Kingston, ON, K7L 3N6, Canada; E-mail: cressler@queensu.ca\\


\vspace{1cm}

\subsubsection*{Abstract}

Phylogenetic comparative analysis is an approach to inferring evolutionary process from a combination of phylogenetic and phenotypic data.
The last few years have seen increasingly sophisticated models employed in the evaluation of more and more detailed evolutionary hypotheses, including adaptive hypotheses with multiple selective optima and hypotheses with rate variation within and across lineages.
The statistical performance of these sophisticated models has received relatively little systematic attention, however.
We conducted an extensive simulation study to quantify the statistical properties of a class of models toward the simpler end of the spectrum that model phenotypic evolution using Ornstein-Uhlenbeck processes.
We focused on identifying where, how, and why these methods break down so that users can apply them with greater understanding of their strengths and weaknesses.
Our analysis identifies three key determinants of performance: a discriminability ratio, a signal-to-noise ratio, and the number of taxa sampled.
Interestingly, we find that model-selection power can be high even in regions that were previously thought to be difficult, such as when tree size is small.
On the other hand, we find that model parameters are in many circumstances difficult to estimate accurately, indicating a relative paucity of information in the data relative to these parameters.
Nevertheless, we note that accurate model selection is often possible when parameters are only weakly identified.
Our results have implications for more sophisticated methods inasmuch as the latter are generalizations of the case we study.

\noindent
\textbf{Keywords:}
phylogenetic comparative analysis, adaptation, model selection,
evolutionary model, Ornstein-Uhlenbeck

\newpage

Central to modern comparative methods are the mathematical models that connect phenotypic evolution to the branching pattern of the phylogeny.
In recent years, creative application of these models has allowed biologists to test complex hypotheses involving adaptive radiation \citep{Setiadi2011b,Monteiro2011,Butler:2004}, multiple rates of evolution \citep{OMeara:2006,Venditti:2011,Beaulieu:2012}, and specialization to environmental \citep{Collar2010b,Collar2011,Edwards2010,Gomez2007,Kozak2010a,Labra2009} or behavioral factors \citep{Collar2009a,Monteiro2011,Scales:2009}.
Genuine progress is made when meaningful alternative hypotheses are compared and some are rejected.
Examples include tests of whether muscle properties of animals evolve in response to the need to flee predators vs.\ the need to chase down prey \citep{Scales:2009} and whether floral evolution is driven by pollinator shifts vs.\ coevolution between flowers and pollinators \citep[i.e., gradual vs.\ punctuated evolution;][]{Whittall2007}.

The ability to formulate and test these hypotheses has been made possible by an expansion of modeling strategies over the last 30 years.
Model-based approaches to phylogenetic comparative methods began with application of the Brownian-motion (BM) model, which captures both the stochastic nature of evolutionary change in phenotypes and the expectation of similarity due to common ancestry \citep{Felsenstein:1973,Felsenstein:1985}.
More recent extensions of BM allow rates of stochastic evolution to vary across the phylogenetic tree \citep{OMeara:2006,Eastman:2011,Revell:2011,Venditti:2011}.
\citet{Lande:1976,Lande:1980} expanded the model repertoire, introducing the Ornstein-Uhlenbeck model as a formal means of modeling both natural selection and genetic drift in macroevolution.  Inspired by \citeauthor{Simpson:1953}'s~\citeyearpar{Simpson:1953} notion that major evolutionary diversification events occur when species enter new ``adaptive zones'', Lande conceptualized phenotypic evolution on a fitness landscape, on which adaptation pulls traits toward one of several optima and stochastic forces introduce variation in trait values but also, importantly, allow switching between adaptive zones.
The OU model's accounting for selection, noise, and multiple trait optima allow it to be used to represent alternative evolutionary histories and thus make it a valuable tool in phylogenetic comparative analysis.
\citet{Hansen:1997} pioneered the use of the OU process with multiple optima in comparative analysis.
Later, \citet{Butler:2004} put Hansen's method into an information-theoretic framework and illustrated how multiple competing evolutionary hypotheses could be tested via the comparison of alternative evolutionary-history scenarios.
More recently, the SLOUCH model expands the range of testable hypotheses still further, allowing hybrid OU-BM models and a form of evolutionary regression between phenotype and covariates \citep{Hansen:2008,Labra2009}.
Both the OUCH and SLOUCH models have multivariate extensions \citep{King:2009,Bartoszek:2012}.
Another recent expansion implements multiple-optimum OU models in which noise intensity and selection strength can vary across lineages \citep{Beaulieu:2012}.

It is important to note that the models just described can be used to achieve various goals.
When the goal is the evaluation of a relatively small set of distinct, biologically informed hypotheses, information-based model selection approaches are useful as a means of evaluating relative explanatory power \citep{Burnham:2002,Butler:2004,OMeara:2006,Beaulieu:2012,Bartoszek:2012}.
Typically, the selection of models is not an end in itself but is rather an aid to reasoning about evolutionary mechanisms and drivers.
In such a context, one bears in mind that the larger the number of models entertained, the less reliable are the inferences obtained.
This is a problem that the use of information criteria only incompletely ameliorates \citep{Boettiger:2012,Ho2014}.
For this reason, inferences are often clearest when the hypotheses are laid out \emph{a priori}.

By contrast, when the goal is to reconstruct ancestral states, one wishes to avoid \emph{a priori} limitation on the range of possible histories.
In particular, one may seek to infer, from knowledge of the pedigrees and traits of extant organisms, ancestral character states and/or historical shifts among selective regimes.
In such a context, one typically must winnow through a large number of alternative histories to identify those most consistent with the data \citep{Hipp:2010,Eastman:2011,Ingram:2013,Revell:2011,Uyeda2014}.
Such an exercise is perhaps best thought of as a form of exploratory data analysis.

Regardless of the ultimate goal, however, greater flexibility and realism in models comes at the cost of more mathematical complexity, more parameters to estimate, and more models to compare.
In effect, as parameter space dimension increases, the information contained in data becomes more dilute, the precision with which parameters can be estimated diminishes, and the risk of overfitting grows \citep{Ho2014}.
We set out to identify the principal determinants of model-selection power and parameter estimation accuracy in a class of models of intermediate complexity by means of an extensive simulation study using OU models with multiple adaptive optima (the OUCH class of models).
The R package \texttt{ouch} implements these methods, but other software implementations exist, for example in the packages \texttt{ape} \citep{Paradis:2004}, \texttt{geiger} \citep{Harmon:2008}, and \texttt{OUwie} \citep{Beaulieu:2012}.
Our results directly concern the reliability of model selection, the accuracy of parameter estimation, and the design of comparative studies based on such models.
In addition, our results have implications for inferences based on more complex models, inasmuch as the latter are extensions of the OUCH class.
The results of this study apply directly to the specific models implemented in the \texttt{ouch} package and with equal force to other software implementations of the same models and, moreover, to more sophisticated models, of which the OUCH models are special cases \citep[e.g.,][]{Hansen:2008,Bartoszek:2012,Beaulieu:2012}.
We detail these implications in the Discussion.

In the following, we first describe and explain the models explored and point out the key quantities informed by data.
Next, we outline the design of our simulation study.
We then present results regarding the determinants of model selection power (probability of choosing the correct model) and parameter estimation accuracy (expected error in estimated parameters).
We identify three dimensionless numbers that effectively determine statistical power and discuss how other features of the data affect inference reliability.
Some of the latter, such as the size and shape of the phylogenetic tree and the distribution of selective regimes on the tree, may be affected by study design via choice of taxa sampled;
we quantify the relative importance of each of these.
Importantly, we find that large errors in parameter estimates are quite common even in circumstances under which power is high.
Finally, we discuss the implications of our results for study design, for the interpretation of analytical results, and for more sophisticated OU-based methods.

\section*{Methods}

\begin{figure}
 \includegraphics[width=0.7\textwidth]{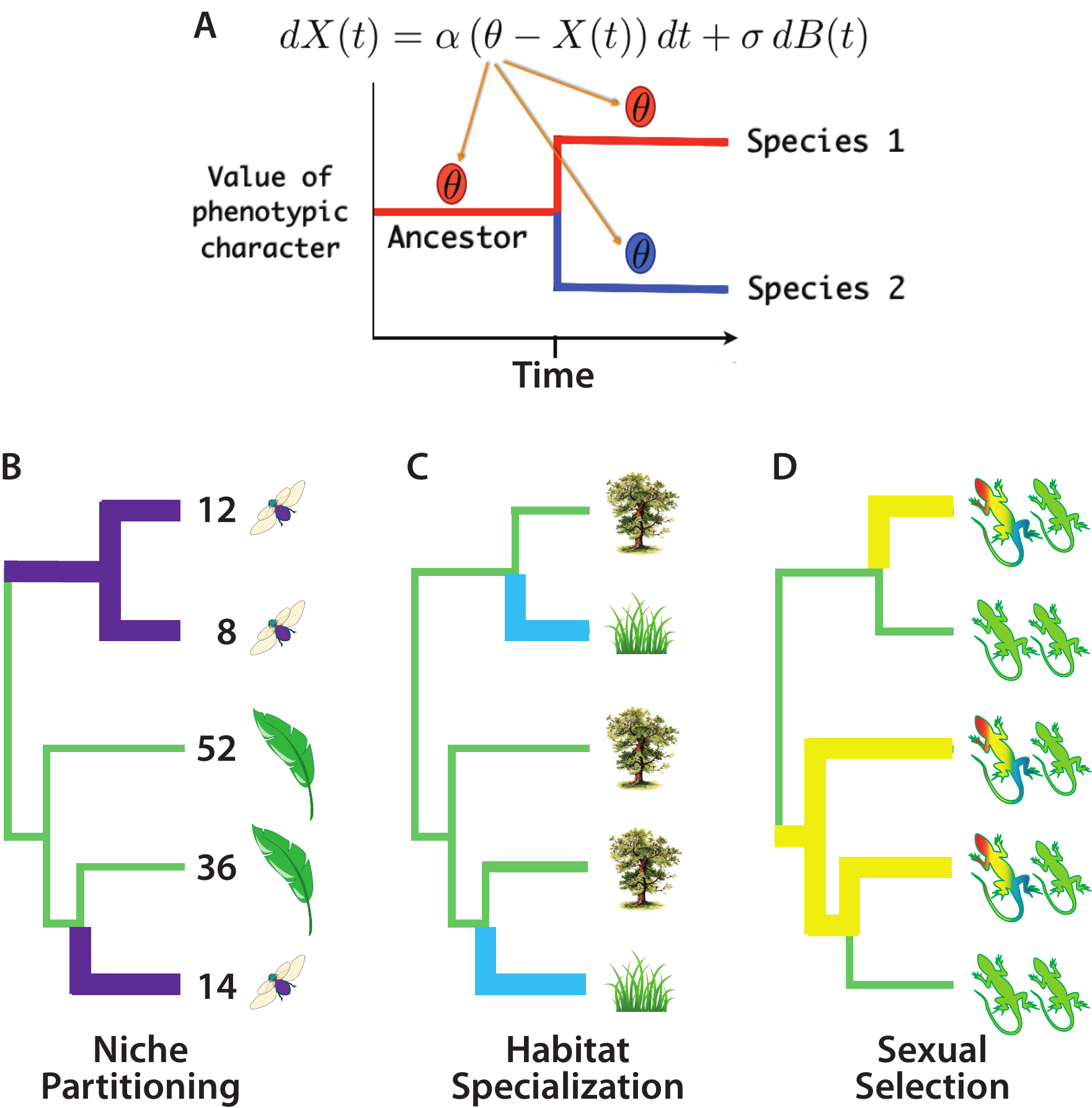}
 \caption{
   (A) \texttt{ouch} models the evolution of quantitative traits as an evolutionary change in the mean phenotypic value, which depends on the selective regimes operating during evolutionary history.
   Given phenotypic data and a phylogenetic tree, investigators can specify evolutionary hypotheses by ``painting'' the branches of the phylogenetic tree with the selective regime assumed to be operational along that branch.
   \texttt{ouch} fits a model with common values of $\alpha$ and $\sigma$, but with different optima on branches as specified by the investigator.
   (B--D) Hypothetical alternative hypotheses for the evolution of body size in a clade of lizards.
   (B) A hypothesis of niche partitioning by food type for the evolution of body size (numerical values).
   Animals specializing on an insect diet enter a new selective regime indicated by purple.
   (C) A hypothesis of specialization to habitat type, with shifts occurring when tree-dwelling lineages enter a grassy habitat.
   (D) A hypothesis of sexual selection driving evolutionary change in body size, with sexual selection indicated by color dimorphism.
   Each of these evolutionary scenarios can be expressed as a separate \texttt{ouch} model.
   Model fitting can then determine which hypothesis is best supported by data.
   \label{fig:hypothesis}
 }
\end{figure}

\subsection*{Models of evolution for quantitative characters}

Both Brownian motion (BM) and Ornstein-Uhlenbeck (OU) models were used in this study.
We provide only a brief account of these models here; both have been described in detail previously \citep{Hansen:1997, Butler:2004}.
The BM model for phenotypic evolution assumes that evolutionary changes along a lineage are stochastic and uncorrelated in time.
In particular, over any time interval, the phenotype changes by a random amount, the magnitude of which has mean zero and variance proportional to the duration of the interval and is independent of changes over any other interval.
Under BM, therefore, observed correlations among the traits of extant taxa are attributed to a combination of chance and shared evolutionary history.
Mathematically, BM is the It\^o diffusion
\begin{equation*}
  \begin{gathered}
    \dd{X(t)} = \sigma\,\dd{B(t)}, \qquad X(0) = X_0,
  \end{gathered}
\end{equation*}
where the phenotypic trait $X(t)$ evolves through time $t$ along each lineage, random deviations are introduced by the Gaussian white noise $\dd{B(t)}$---usefully visualized as a normal random variable with mean zero and variance $\dd{t}$---and the magnitude of these deviations depends on the noise intensity, $\sigma$.
Previous authors, including ourselves, have used the term ``drift'' in connection with this noise.
It is important to note that stochasticity may arise not only from genetic drift but from other, unmodeled factors.
Here and in the following, we adopt the convention that $t=0$ at the root of the phylogeny.
When BM is fit to data in a comparative analysis, the only parameters to be estimated are $\sigma$ and the phenotype at the phylogeny's root, $X_0$.

The OU model expands on BM by including a deterministic component that models the tendency of a trait to evolve towards an adaptive optimum.
In this model, this tendency is scaled by the parameter $\alpha$, which quantifies the rate at which the trait is pulled towards an optimum, while the optimum trait values, which may change with time and across branches of the phylogeny, are denoted $\theta(t)$.
From the point of view of \citet{Simpson:1953} and \citet{Lande:1976,Lande:1980}, $\alpha$ is interpreted as ``strength of selection''.
Formally, the OU process is an It\^o diffusion satisfying
\begin{equation}\label{eq:OU}
  \dd{X(t)}=\alpha\,\big(\theta(t)-X(t)\big)\,\dd{t}+\sigma\,\dd{B(t)}.
\end{equation}

\subsection*{Additional assumptions}

To define an OU process, one must make some assumption about the root-state, $X(0)$.
One may follow \citet{Hansen:1997} and \citet{Butler:2004} in introducing a parameter, $X_0=X(0)$, as in the BM model.
This assumption implies that BM is the special case of the OU model for which $\alpha=0$.
It frequently happens, however, that the data carry little information on $X_0$.
In fact, \citet{Ane:2008} proved that under Brownian motion evolution, the estimate of the root-state is inconsistent, tending to a random value as sample size increases.
Moreover, for OU models, one sometimes finds that estimates of $X_0$ and $\alpha$ are correlated.
The alternative we adopt is to follow \citet{Hansen:2008} and \citet{Ho2013} in assuming that $X(0)$ is distributed according to the stationary distribution of an OU process:
\begin{equation}\label{eq:OU-stat}
  X(0)\;\sim\;\mathrm{normal}\left(\theta(0),\frac{\sigma^2}{2\,\alpha}\right).
\end{equation}
[The stationary trait distribution about a particular optimum is normal with mean $\theta$ and variance $\frac{\sigma^2}{2\,\alpha}$.]
It is important to note that, under this definition, BM is no longer nested within OU.
Nor can we redefine BM to make it so, since BM lacks a stationary distribution.

In the foregoing, the time-dependence of the optimum, $\theta$, has been left general.
The OUCH class of models restricts this dependence to a particular form: $\theta(t)$ is piecewise constant on intervals and takes values in a finite set $\{\theta_i\}$, the set of ``selective regimes'' or just ``regimes'' (a term we prefer to Simpson's ``adaptive zones'').
Moreover, it is assumed that the intervals and regimes are known, though the optimum trait values associated with each regime are not.
As an aid to visualization, one can associate a color with each selective regime.
Applying the corresponding color to each interval on the phylogeny, one obtains a ``painting'' of the phylogeny (Fig.~\ref{fig:hypothesis}).
The OUCH class further assumes that $\alpha$ and $\sigma$ are constants.
A frequent misconception about this class of models is that regime shifts must co-occur with phylogenetic branch points within this class of models.
This is incorrect: one can work around a software implementation that allows regime shifts only at nodes by inserting additional nodes as needed.

\subsection*{Dimensionless parameter combinations}

The fact that the variables and parameters of the OU model have units presents a challenge to the design of a simulation study.
In Eq.~\ref{eq:OU} in particular, $X$ and $\theta$ share the same units, which depend on the trait in question, the tree depth $T$ has the same units as $t$, $\alpha$ has units of inverse time, $\dd{B}$ has units of time to the $\tfrac{1}{2}$ power, and $\sigma$, therefore, has units of trait times time to the $-\tfrac{1}{2}$ power.
The statistical performance of any model, however, cannot depend on the arbitrary choice of units.
Thus, while a change in the scales used to measure $t$ or $X$ will change the numerical values of $\alpha$, $\sigma$, and $\theta_i$, statistical inference must depend only on dimensionless combinations of these parameters \citep{Buckingham1914}.
We can identify these dimensionless combinations via the simple mathematical method of non-dimensionalization, which involves standardizing the problem by rescaling model variables.
Non-dimensionalization reduces the number of parameters to the minimal set needed to produce the full range of model behaviors.
While there is typically some freedom in the choice of scaling, judicious choice results in dimensionless parameters that have meaningful interpretation (for example, for swimming animals, non-dimensionalizing the Navier-Stokes equations of fluid dynamics by animal length and speed leads to the Reynolds number, which determines whether flow is laminar or turbulent).
\citet{Strang1986} and \citet{Murray2002} give good expositions of non-dimensionalization.
Here, we derive three dimensionless quantities that, as we will show, largely determine the statistical power of OU-based model selection and the accuracy of parameter estimates.

We proceed by identifying natural scales on which to measure the variables.
An obvious choice of time-scale is the total depth, $T$, of the phylogeny.
Accordingly, we define the dimensionless time $\widetilde{t}={t}/{T}$.
Several choices of scaling for $X$ are available.
In earlier studies, we and others have used an arbitrary scaling for $X$.
Alternatively, letting $\dlta{\theta}$ denote the smallest difference between two selective regimes, one can define the dimensionless trait $\widetilde{X}={X}/{\dlta{\theta}}$ and dimensionless optimum $\widetilde{\theta}(\widetilde{t})=\theta(t/T)/{\dlta{\theta}}$.
Then the OU process for lineage $i$ obeys
\begin{equation*}
\dd{\widetilde{X}_i}=(\alpha\,T)\,\left[\widetilde{\theta}_i(\widetilde{t})-\widetilde{X}_i\right]\,\dd{\widetilde{t}}+\left(\frac{\sigma\,T^{1/2}}{\dlta{\theta}}\right)\,\dd{\widetilde{B}_i}=\eta\,\left[\widetilde{\theta}_i(\widetilde{t})-\widetilde{X}_i\right]\,\dd{\widetilde{t}}+\gamma\,\dd{\widetilde{B}_i},
\end{equation*}
where $\dd{\widetilde{B}}$ is an increment of a dimensionless Brownian motion process.
The dimensionless parameter $\eta=\alpha\,T$ can be viewed as a measure of the depth of the phylogenetic tree relative to the characteristic timescale associated with adaptation.
Herein, we refer to $\eta$ as the \emph{selection opportunity}, since it involves both the strength, $\alpha$, of selection and the amount of time it has had to work.
It is proportional to the depth of the phylogeny measured relative to the phylogenetic half-life $\frac{\log{2}}{\alpha}$ \citep{Hansen:1997}.
Note that $\eta$ is not to be confused with the population-genetic concept of opportunity for selection \citep{Arnold1984}.
The quantity $\gamma=\frac{\sigma\,T^{1/2}}{\dlta{\theta}}$ is a dimensionless measure of the noise intensity.

Two particular combinations of $\eta$ and $\gamma$ will turn out to be important from the point of view of statistical analysis.
The first, $\phi=\frac{\sqrt{2\,\eta}}{\gamma}=\frac{\sqrt{2\,\alpha}\,\dlta{\theta}}{\sigma}$---which we will call the \emph{discriminability ratio}---measures the smallest difference between selective optima relative to the standard deviation of the stationary trait distribution.
When $\phi$ is large, the optima are distinct relative to chance variability around them; when $\phi$ is small, chance variability will obscure the difference between the closest optima, even when selection has been acting over very long times.
The second quantity, $\sqrt{\eta}\,\phi$, combines the selection opportunity and the discriminability ratio.
Rewriting, we see that $\sqrt{\eta}\,\phi=\frac{\sqrt{2\,T}\,\alpha\,\dlta{\theta}}{\sigma}$ is proportional to the ratio of $\alpha$, which quantifies the strength of the deterministic signal in Eq.~\ref{eq:OU}, to the noise intensity, $\sigma$.
For this reason, we refer to $\sqrt{\eta}\,\phi$ as the \emph{signal-to-noise ratio} (SNR).
It will emerge that the SNR is a predictor of statistical power.

It is worth noting at this point that, while the particular non-dimensionalization we have chosen leads to insight into the statistical properties of OU-based comparative methods, it does not immediately translate into a prescription for data analysis, i.e., it is not straightforward to fit this dimensionless model to data.
We return to this issue in the Discussion.

\subsection*{Plan of the simulation study}

The simulation study was designed to investigate the influence of aspects of the data-generating process on model-selection power and parameter-estimation accuracy.
We varied the number of taxa sampled, the values of $\alpha$ and $\sigma$, and the arrangement (``painting'') of selective regimes over the phylogeny.
Since tree height, $T$ and minimum inter-optimum distance $\Delta\theta$ set the scales of time and phenotype, respectively, we fixed $T=1$ and $\Delta\theta=1$ without any loss of generality.
We chose $(\eta,\phi,\text{tree size})$ triples over a broad range of parameters using a Latin hypersquare design (see below and Fig.~\ref{fig:paramcombs}).
We simulated phylogenetic trees, regime paintings, and phenotypic data under each parameter set and fit each of six competing models to the data as described below.
We assessed model-selection power as the probability that the true model had the lowest $\mathrm{AIC}_c$ score.
In addition, when the true model was selected, we assessed parameter-estimation accuracy as mean squared error (MSE) in each of the estimated parameters.
We investigated how power and accuracy depend upon the selection opportunity, discriminability ratio, tree size, tree balance, and evenness of regime representation across both the tips and the branches of the phylogeny.

Fig.~\ref{fig:outline} gives a schematic overview of the study design.
In all, the simulation study used
220 parameter triples,
40 sets of regime paintings per parameter triple,
40 trees of each size per regime painting,
and 40 sets of simulated phenotypic data per tree.
In all, then, we fit each of the six models to \scinot{14.08}{6} simulated datasets, using about \scinot{7}{3}~cpu-hr of computation.
Further details of the simulation study are given below.
Computations were carried out in the R statistical computing environment \citep{R}.
We next describe the simulation study in more detail.
All codes necessary to fully reproduce this study in all particulars are provided in the Supplement.

\subsubsection*{Generating random phylogenies}

Because the raw data for phylogenetic comparative analyses are typically hard-won and statistical performance is most an issue when sample sizes are small, we focused our investigation on phylogenies of between 10 and 50 extant species, a modest number representative of many studies in the literature.
We encountered no reason to doubt that larger sample sizes result in better inferences generally.
Ideally, we would simulate phylogenies from a model known to generate realistic trees.
However, some have questioned the ability of generic tree models, such as pure birth or birth-death processes, to generate trees with shape properties similar to empirical phylogenies \citep{Mooers1997}.
More generally, we know of no generative model for realistic phylogenies.
We therefore randomly subsampled a large empirical ultrametric phylogeny of 187 species of Caribbean \emph{Anolis} lizards \citep{Nicholson:2005kn}.
Although phylogenies so generated may or may not be \emph{realistic}, they are indubitably \emph{real}.
Note too that this method of generating phylogenies bears some resemblance to that in which real phylogenies are obtained, viz., by subsampling the Tree of Life.
Be this as it may, the phylogenies so generated display a wide range of topologies and shapes.
Fig.~\ref{fig:treemetrics} shows the distribution of several metrics of tree shape \citep{Mooers1997} for the 65600 distinct trees used in this study.

\subsubsection*{Selecting model parameters}

In order to develop an understanding of where, how, and why statistical performance of the OUCH-class methods varies, we focused on the region of parameter space where statistical power and parameter-estimation accuracy range from very high to quite low.
In particular, we varied the dimensionless parameters $\eta$ and $\phi$ over the range $(0.2,5)$, which represents a broad spectrum of scenarios, from ones in which the data contain very little information about the evolutionary process to ones where recovery of the true model is nearly guaranteed.
As $\eta$ varies from $0.2$ to $5$, the depth of the tree as measured in phylogenetic half-lives grows from $0.29$ to $7.2$.
As $\phi$ varies over the same range, the separation of adjacent optima grows from $1/5$ to $5$ standard deviations of the stationary distribution.
As discussed above, we varied tree size (number of sampled taxa) from 10 to 50.
Specifically, we used a Sobol' low-discrepancy sequence \citep{Press1992} to generate 220 $(\eta,\phi,\text{tree size})$ triples.
We computed $\alpha$ and $\sigma$ according to the relations described above.
Fig.~\ref{fig:paramcombs} shows the range of parameters used in the study.

\subsubsection*{Randomizing over disposition of selective regimes}

40 sets of 3-regime paintings were simulated on the large \emph{Anolis} phylogeny by assuming that regimes shifts occur according to a Poisson process along each lineage and that shifts occur randomly among the regimes.
Specifically, the expected waiting time between shifts was $0.6\,T$ and the identity of each new regime was selected randomly with equal probability from among the two candidates.
The 40 regime paintings are shown in the Supplement.
Each sampled subtree inherited its painting from the large phylogeny.

\subsubsection*{Computing tree and regime metrics}

Tree shape and the distribution of regimes are to some extent a function of experimental design, as they are determined by the choice of taxa included in a study.
We therefore aimed to quantify their effects on statistical performance.
For each subtree, we calculated several measures of tree shape and regime distribution.
To quantify tree shape, we required a measure able to accommodate polytomies.
We chose a measure based on \citeauthor{Sackin1972}'s \citeyearpar{Sackin1972} index, as follows.
For each tip $i=1,\dots,n$, let $N_i$ be the number of nodes between the tip and the root (including the root).
The average, $\bar{N}=\tfrac{1}{n}\,\sum_i\!N_i$, over all tips yields a measure of imbalance.
We normalized this index by dividing by its expectation, $\Expect{\bar{N}}=2\,\sum_{i=2}^n\!\tfrac{1}{i}$, under a coalescent model \citep{Kirkpatrick1993}, which yields a quantity roughly independent of tree size.
We refer to this metric as \emph{tree imbalance} since larger values indicate a more unbalanced tree.

We also assessed how the distribution of selective regimes among tips and along lineages affects statistical performance.
To calculate tip evenness, $J_{T}$, we used the standardized Shannon diversity index
\begin{equation*}
  J_{T}=\frac{-\sum_{i=1}^S\!p_i\log{p_i}}{\log{S}},
\end{equation*}
where $S=3$ is the number of regimes and $p_i$ is the fraction of tips in each regime.
Evenness of regime representation along the branches, $J_{B}$, was computed by applying the same Shannon diversity index formula to the proportion of total branch-length in each regime.
Both $J_T$ and $J_B$ vary between 0 and 1; higher values correspond to more even distribution of regimes.
As Fig.~\ref{fig:treemetrics} shows, no strong correlations between tree size and these metrics were in evidence among the trees we used.

\subsubsection*{Simulating the phenotypic data}

For each parameter combination and regime painting of the large tree, we generated 40 phenotypic datasets using the stochastic 3-regime OUCH model.
We assumed equal distances between adjacent optima.
With this assumption, location of the optima is inconsequential once the scale of $X$ has been chosen;
we therefore set $\theta_A=-1$ without additional loss of generality, and $\theta_B=0$, $\theta_C=1$.

\subsubsection*{Model fitting}

To evaluate model-selection power, we fit a panel of models to the simulated data, assuming known phylogeny and regime painting, using the \texttt{ouch} R package \citep{King:2009}.
In particular, we considered Brownian motion (BM), a single-optimum OU model (OU1), two 2-regime OU models (OU2ab, OU2bc), the true model (OU3), and a fully non-phylogenetic model (NP).
The two 2-regime OU models are nested within the OU3.
In the OU2ab, the two large-value regimes (B \& C) are collapsed into a single regime;
in the OU2bc, regimes A \& B are collapsed into one.
The NP model is equivalent to the ``white noise'' model implemented in \texttt{geiger} \citep{Harmon2008,Hunt2006}.
Under it, each species' phenotype is modeled as an independent sample from a normal distribution with mean depending on the tip regime for that species and common variance.

Maximum-likelihood model fitting was performed using a combination of Nelder-Mead and subplex optimization algorithms \citep{Rowan:1990,subplex}.
We performed model selection using the Akaike Information Criterion corrected for small sample size \citep[$\mathrm{AIC}_c$;][]{Burnham:2002}.
The best model was deemed simply to be that with the lowest $\mathrm{AIC}_c$; this ignores the potential for ambiguous or inconclusive model-selection results which arise in practice when one or more models have $\mathrm{AIC}_c$ scores differing by a few units only.
\citet{Boettiger:2012} offer a convincing demonstration that over-reliance on information criteria can lead to unreliable conclusions.
They very sensibly advocate a case-by-case approach in which the parametric bootstrap is used to compute more accurate P-values for information-criterion-based model selection.
Obviously, such a case-by-case, bootstrap-based approach would be too computationally expensive for our purposes.
However, we employ this bootstrap analysis below in the examination of a few illustrative cases.
Furthermore, we anticipate that the breakdown of our raw-$\mathrm{AIC}_c$ approach will be most common at the frontiers dividing high-power from low-power parameter regions and that our principal conclusions will be robust to the choice of model-selection criterion.

We note that our assumptions of known phylogeny and regime painting ignore potentially large sources of uncertainty and therefore make our estimates of statistical performance optimistic: power and accuracy are almost certainly overestimated.
However, as assessment of power was a primary goal of the study, it was essential that the true regime painting be included among the competing hypotheses.
Moreover, since we initiated the iterative optimization algorithms at the true parameter values, our results likely understate the effects of estimation error, a sometimes nontrivial issue in actual data analysis.
Again the effect is to make our assessment of statistical performance optimistic.

\subsubsection*{Assessing power and accuracy}

Power was calculated at each parameter-value/phylogeny/painting combination as
\begin{equation}\label{eq:power}
  \text{power}=\frac{R+\frac{1}{2}}{N+1},
\end{equation}
where $N$ is the number (40) of replicate phenotypic datasets and $R$ is the number for which the true evolutionary model was better supported than the alternatives.
Eq.~\ref{eq:power} is slightly more stable than $R/N$ and disallows the possibility of power $=0$ or $1$.

We assessed the accuracy of parameter estimates when the true model was selected using mean squared error (MSE):
\begin{equation*}\label{eq:MSE}
  \text{MSE}(\hat{\beta})=\frac{1}{R}\,\sum_i\left(\hat{\beta_i}-\beta\right)^2,
\end{equation*}
where $\beta$ stands for the true value of any one of the model parameters, $\hat{\beta_i}$ is the estimate in the $i$th replicate, and the sum is taken over the $R$ replicates for which the true model was best supported.
In restricting attention to parameter estimates from the best-supported model only, we reasoned that an investigator using this method would be uninterested in parameter estimates from unsupported models.

We explored the dependence of power and MSE on $\eta$, $\phi$, tree size, tree imbalance, and the two regime-evenness metrics, tip evenness, $J_T$, and branch evenness, $J_B$.
We quantitatively assessed these relationships using generalized additive model (GAM) regression of power on these predictor variables \citep[][via the \texttt{mgcv} R package]{Wood2006}.
Summaries of the best-fitting GAMs and several alternatives are given in Appendix~\ref{sec:gams}.

\section*{Results}

\subsection*{Model selection power}

\begin{figure}
\includegraphics[width=\linewidth]{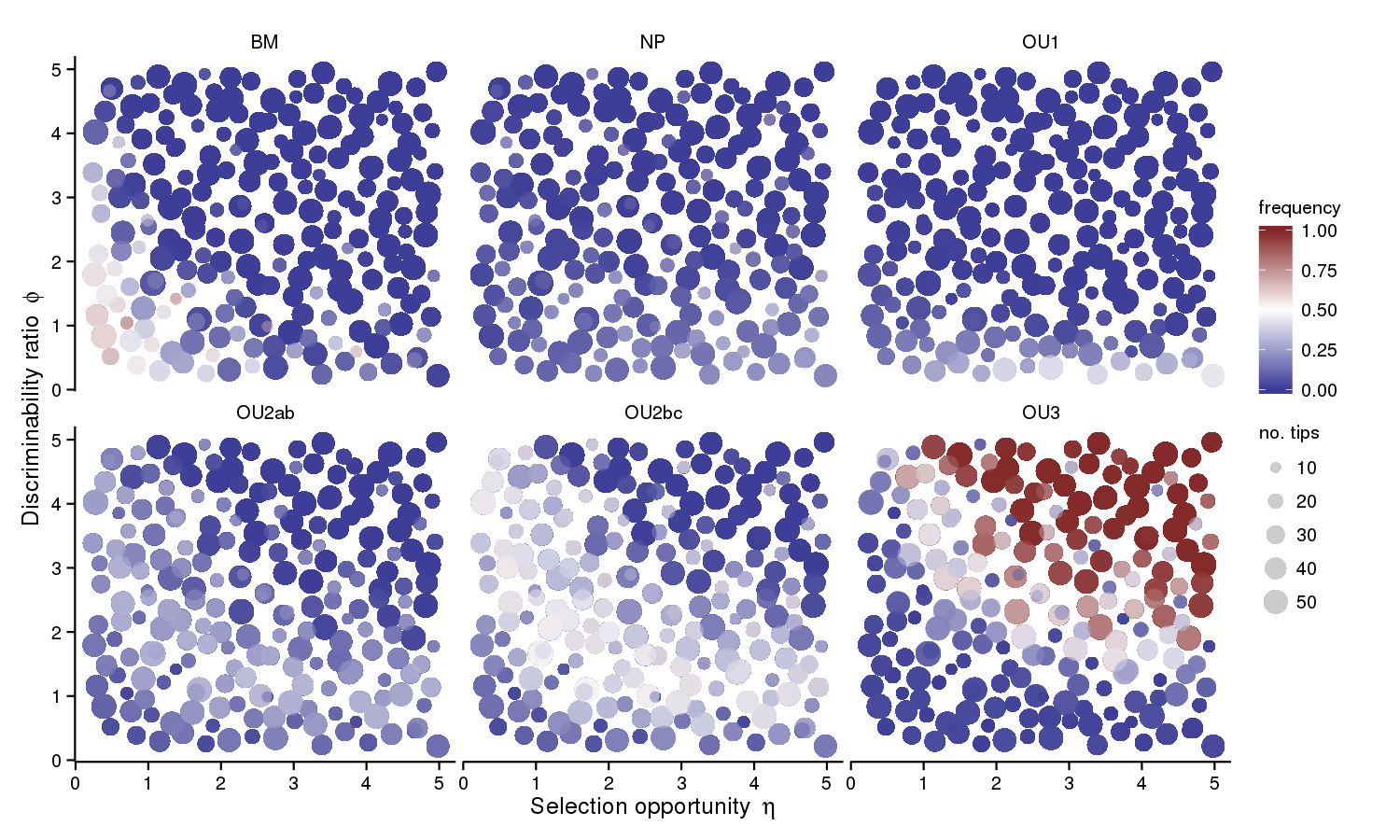} \hfill{}
 \caption{
   The frequency (fraction of 40 phenotypic datasets) that each of the alternative models was best-supported as a function of the dimensionless simulation study variables: selection opportunity $\eta$, discriminability ratio $\phi$, and tree size.
 }
  \label{fig:modelprobs}
\end{figure}

Overall, the simulation study encompassed regions of high, low, and moderate power.
In some regions of parameter space, power approached 100 percent, especially when selection opportunity $\eta$ and discriminability ratio $\phi$ were high, and 20 or more taxa were included (Fig.~\ref{fig:modelprobs}).
Other regions of parameter space had very low power, such that the true model was rarely selected.
The GAM regressions indicated that the strongest predictors of power were $\eta$, $\phi$, tree size, and the evenness of regimes across the branches.
Unexpectedly, the effect sizes for tip evenness and tree imbalance were both very small and statistically significant only because of the large sample size.
In other words, evenness in the proportion of evolutionary time spent in the various regimes was more predictive of power than were balance in the shape of the tree or evenness in the the number of tip taxa in each regime.

Unsurprisingly, the support for each of the alternative (incorrect) models varied strongly with selection opportunity $\eta$ and discriminability ratio $\phi$ (Fig.~\ref{fig:modelprobs}).
Support for Brownian motion models was high only when selection opportunity $\eta$ was very small.
The OU1 model had highest support when selection opportunity was large but the discriminability ratio $\phi$ was very small, i.e., when selection was strong but the noise was nevertheless strong enough to obscure differences among the optima.
The two alternative OU2 models found high support at moderate values of $\eta$ and $\phi$, whereas the true OU3 model had high support for larger values of $\eta$ and $\phi$.

\begin{figure}
\includegraphics[width=\linewidth]{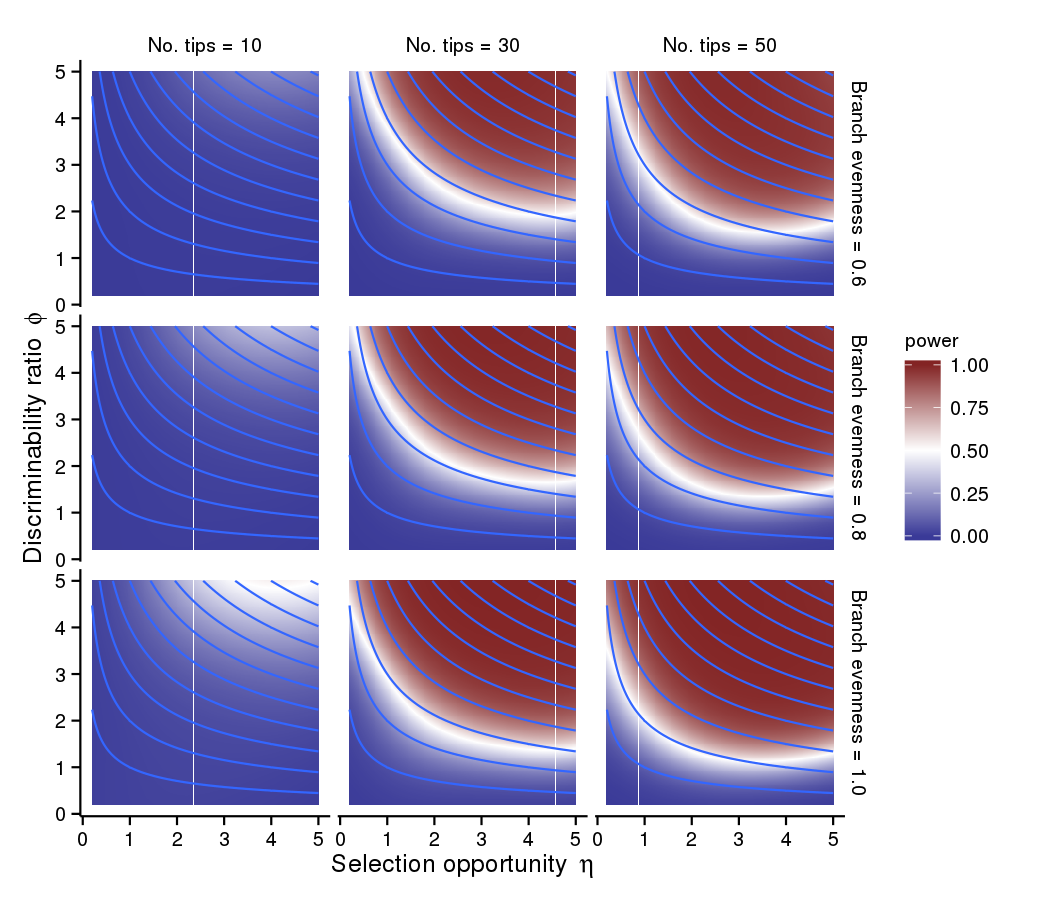} \hfill{}
\caption{
  Power as $\eta$ and $\phi$ are varied for a range of branch evenness ($J_B$) and tree size values.
  Red corresponds to high ($>$50\%) probability of selecting the true model (blue $<$50\%), smoothed using GAM regression, assuming perfectly even tips and a highly balanced tree (imbalance metric of -0.5).
  Fig.~\ref{fig:eta-phi-power-contours-appendix} in Appendix~\ref{sec:supplementaryplots} shows that power is little affected by the tip evenness and tree imbalance.
  Contours show the levels of $\text{SNR}=\sqrt{\eta}\,\phi$.
}
\label{fig:eta-phi-power-contours}
\end{figure}

The regions where alternative models have high support (Fig.~\ref{fig:modelprobs}) have a particular shape, suggesting that some function of $\eta$ and $\phi$ will be informative for power.
Figs.~\ref{fig:eta-phi-power-contours} and \ref{fig:snr-power-contours} show that, to a good approximation, the dimensionless signal-to-noise ratio (SNR), $\sqrt{\eta}~\phi$, together with tree size, is such a determinant, especially when $\phi$ is large.
Fig.~\ref{fig:eta-phi-power-contours} shows how power varies with $\eta$ and $\phi$ for different combinations of tree size and branch evenness.
The correspondence between the contours of power and those of SNR in Fig.~\ref{fig:eta-phi-power-contours} indicates that SNR is a useful predictor of power.
Fig.~\ref{fig:snr-power-contours} shows power as a function of SNR and tree size for a range of branch ($J_B$) and tip ($J_T$) evenness scores.
Tree size plays an important role in determining power:
for very small trees, power is low except at high values of SNR.
This figure also emphasizes the relatively small contribution regime evenness makes to determining power.  
SNR is not a perfect determinant: using $\eta$ and $\phi$ together independently as predictors gives a better GAM fit to the data than using SNR alone, but the improvement in $R^2$ is slight (0.939 vs.\ 0.937, Appendix~\ref{sec:gams}).  

\subsection*{Parameter estimation accuracy}

\begin{figure}
\includegraphics[width=\linewidth]{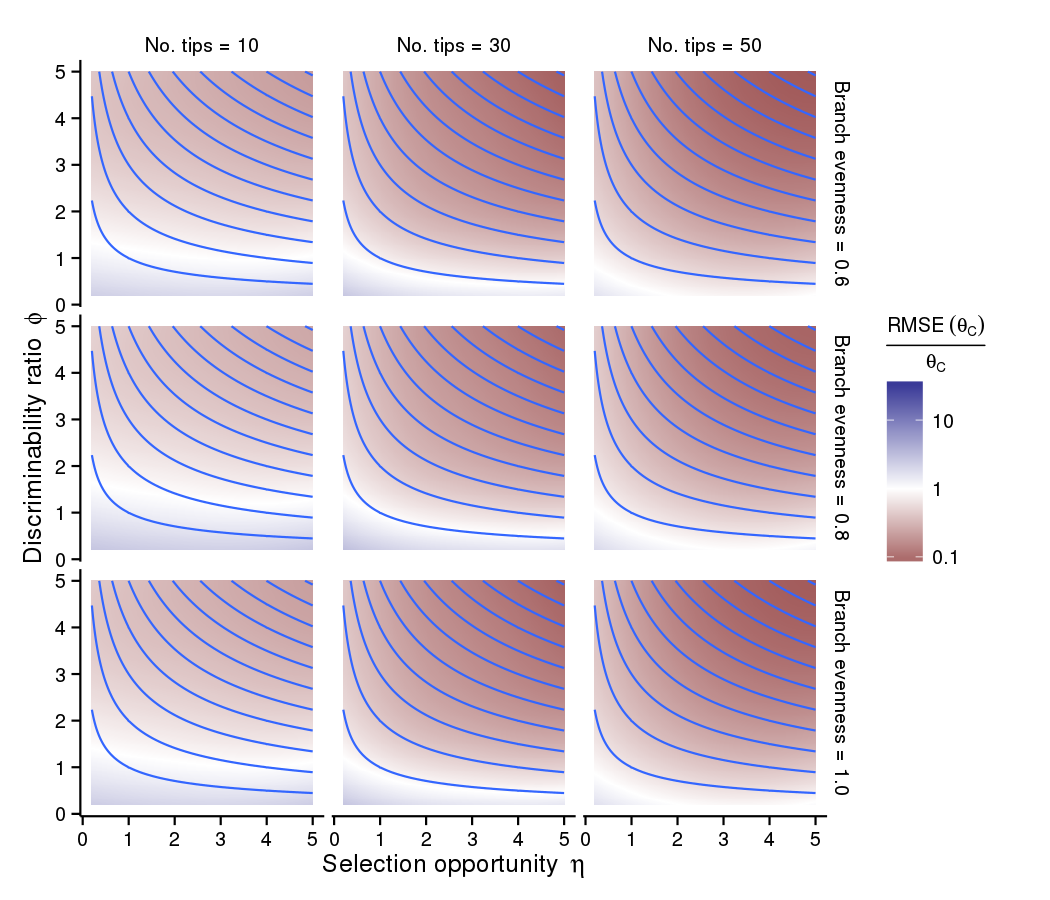} \hfill{}
\caption{
  The relative root mean square error (RMSE) in the estimate of the selective optimum $\theta_C$ as $\eta$ and $\phi$ are varied for a range of tree size and branch evenness values.
  Red hues correspond to less than 100\% relative error;
  blue, to greater than 100\%.
  Predictions are based on the best-fitting GAM regression.
  Contours show the levels of SNR.
}
\label{fig:eta-phi-thetaC-contours}
\end{figure}

\begin{figure}
\includegraphics[width=\linewidth]{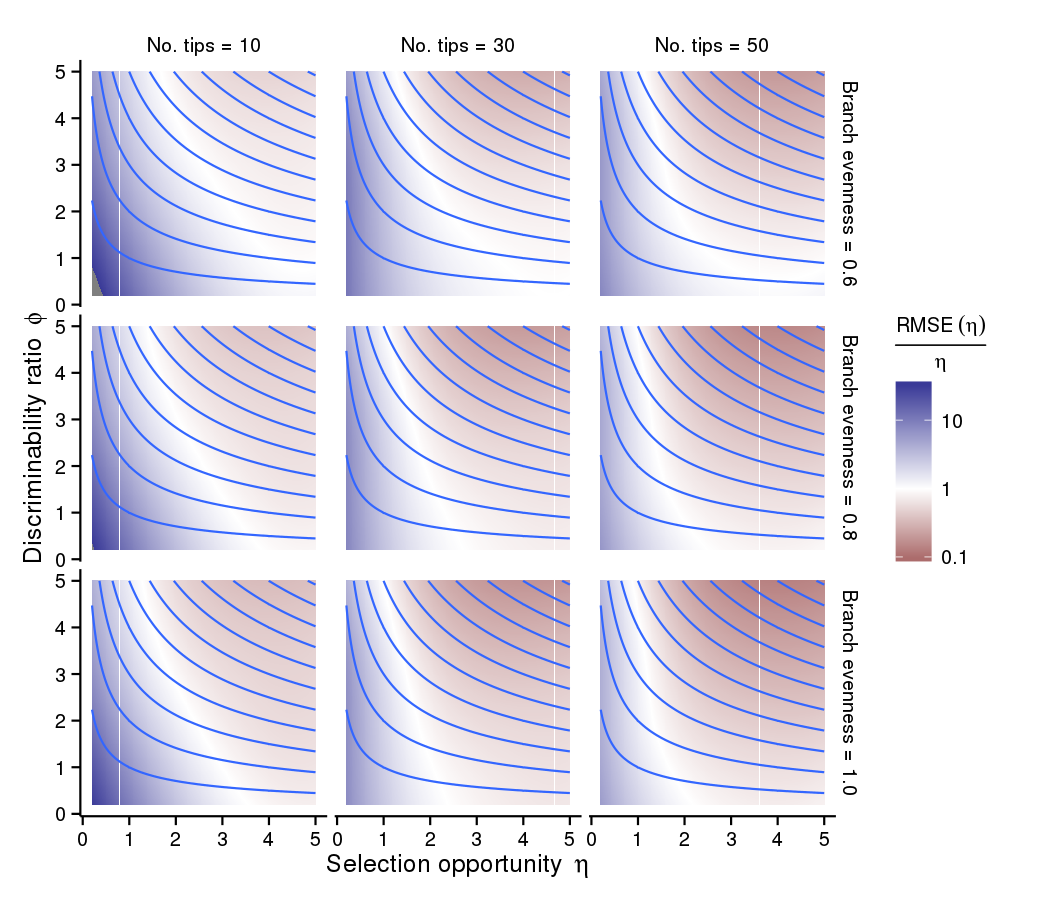} \hfill{}
\caption{
  Relative RMSE in the estimate of selection opportunity ($\hat{\eta}$) as $\eta$ and $\phi$ are varied for a range of tree size and branch evenness values.
  Red hues correspond to less than 100\% relative error;
  blue, to greater than 100\%.
  Contours show increasing SNR.
  Predictions are based on the best-fitting GAM regression.
}
\label{fig:eta-phi-phi-contours}
\end{figure}

\begin{figure}
\includegraphics[width=\linewidth]{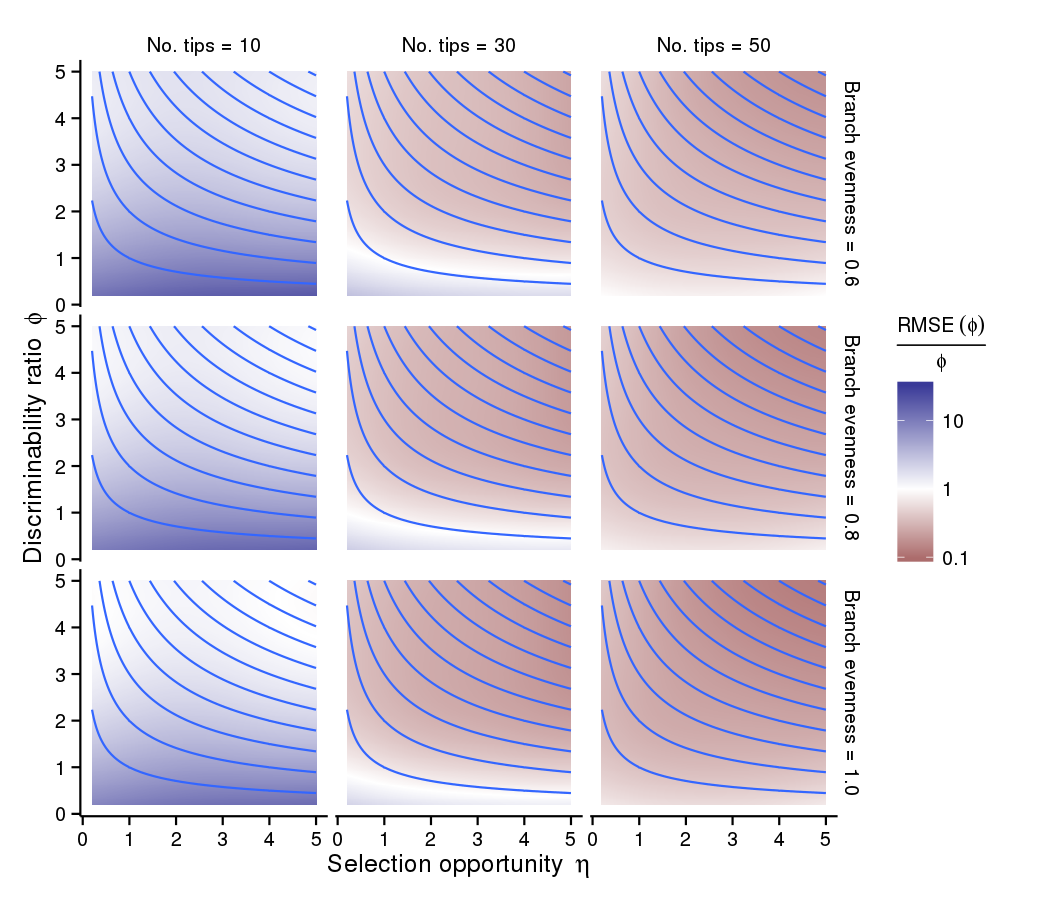} \hfill{}
\caption{
  Relative RMSE in the estimate of discriminability ratio ($\phi$) as $\eta$ and $\phi$ are varied for a range of tree size and branch evenness values.
  Red hues correspond to less than 100\% relative error;
  blue, to greater than 100\%.
  Contours show the levels of SNR.
  Predictions are based on the best-fitting GAM regression.
}
\label{fig:eta-phi-eta-contours}
\end{figure}

The GAM regressions indicated that, as was the case for model-selection power, $\eta$, $\phi$, tree size, and branch evenness were the largest determinants of parameter estimation accuracy.
Figs.~\ref{fig:eta-phi-thetaC-contours}--\ref{fig:eta-phi-eta-contours} depict parameter estimation error based on the best-fitting GAM regressions.
Accuracy of the estimates $\hat{\theta_c}$, $\hat{\phi}$, and $\hat{\eta}$ are presented in terms of relative root mean square error (RMSE).
Errors in estimates of the other two optima (Figs.~\ref{fig:eta-phi-thetaA-contours} and \ref{fig:eta-phi-thetaB-contours}) show a pattern similar to that in Fig.~\ref{fig:eta-phi-thetaC-contours}.
Accuracy of estimates of the optima display a dependence on the selective regime at the root which we describe in the Discussion.

Several important patterns are evident in Figs.~\ref{fig:eta-phi-thetaC-contours}--\ref{fig:eta-phi-eta-contours}.
Of these parameters, it is clear that the selective optima are the best estimated.
Relative errors of 100\% or more are only present for very small values of $\phi$ and are rare when the number of tips is large.
The discriminability ratio $\phi$ is also relatively well estimated, provided the tree is of sufficient size and $\phi$ is not very small.
Selection opportunity is substantially more difficult to estimate accurately:
when either $\eta$ or $\phi$ is small, relative errors exceeding 100\% are common, even when the correct model has been selected.

\section*{Discussion}

This study investigated the statistical properties of methods for phylogenetic comparative analysis based on models of the OUCH class.
Previous studies have suggested that OU-based methods work reliably when sample sizes are large and signal is strong \citep[i.e., many taxa sampled and large SNR, e.g.,][]{Beaulieu:2012,Ho2014}.
This study explored more problematic regions of parameter space with the goal of identifying where, how, and why these methods break down as sample size or SNR become small so that users can apply them with greater understanding of their strengths and weaknesses.

Our results show that model-selection power can be high, even when sample sizes are small, or when the magnitude of the deterministic pull towards the optimum is small.
The critical determinants of power in our analysis are what we have termed the signal-to-noise ratio and the discriminability ratio, both of which involve the distance between optima relative to the stationary variance of the OU process.
We found that parameters can in some circumstances be well estimated, but that care should be taken in interpreting estimates, as parameters can be poorly identified even when model selection is robust.
We discuss these issues further below.

The utility of OU-based methods for comparative analysis is indicated by the number of extensions that have been implemented in recent years.
This includes methods that allow the optima themselves to evolve along the phylogeny according to a Brownian motion process, obviating the need to specify regimes on the phylogeny \citep[SLOUCH,][]{Hansen:2008},
methods that allow for multivariate traits \citep{King:2009,Bartoszek:2012},
methods that allow $\alpha$ and/or $\sigma$ to vary across the phylogeny \citep[\texttt{OUwie},][]{Beaulieu:2012},
and methods that attempt to reconstruct the evolutionary history of regime changes \citep{Hipp:2010,Ingram:2013}.
The conclusions of our study apply with equal force to all implementations of the OUCH class of models and have implications for methods based on generalizations of the OUCH class.

\subsection*{Distinguishing among competing hypotheses}

Our results show that, when selection opportunity, $\eta$, and discriminability ratio, $\phi$, are both sufficiently great and the tree is large enough, the signal of adaptation is strong and we reliably recover the true model (Figs.~\ref{fig:modelprobs}, \ref{fig:eta-phi-power-contours}, and \ref{fig:snr-power-contours}).
It is important to note, however, that when the separation of the optima is small relative to the fluctuations in the evolutionary process (small $\phi$), it is difficult to infer the model accurately, regardless of the value of selection opportunity, $\eta=\alpha\,T$ (Fig.~\ref{fig:eta-phi-power-contours}).
On the other hand, if $\phi$ is sufficiently large, power can be high even when $\eta$ is small.
In the latter situation, to a good approximation, tree size and the signal-to-noise ratio, $\sqrt{\eta}~\phi$, predict model selection power (Fig.~\ref{fig:snr-power-contours}).

In general, it was only when $\eta$ and $\phi$ were both sufficiently large that the OU3 model was reliably recovered.
Certain other regions of parameter space produced data frequently mistaken as having been generated by one of the simpler alternative models (Fig.~\ref{fig:modelprobs}).
Brownian motion found support only when $\eta$ was quite small;
the single-optimum OU model, only when $\phi$ was small.
The two-optimum OU models found some support within a band of $\eta$-$\phi$ combinations for which the phenotypic distributions of species in neighboring selective regimes were likely to overlap, either because selection opportunity was low but discriminability ratio high, or because selection was high but discriminability low.
 Over the range of parameters we examined, the non-phylogenetic model had very little support.
We anticipate, however, that at higher levels of $\eta$ than were explored here, the non-phylogenetic model would find more support.
This is because when $\eta$ is so large that adaptation is essentially instantaneous, the evolutionary history of each lineage becomes irrelevant.
In this situation, the non-phylogenetic model and the true OU model will each fit the data equally well;
the non-phylogenetic model, being simpler, will be preferred by AIC.
In this region, the preference of AIC for the NP model is not evidence against the adaptive hypothesis.
Thus, large values of $\eta$ do not pose challenges to the detection of adaptive evolution.

Finally, although investigators with small datasets may find that increasing sample size rapidly improves power, hope remains for studies based on small numbers of species (Fig.~\ref{fig:snr-power-contours}).
We found that power can be high when SNR is large even with as few as 10 species.
Similarly, \citet{Ho2014} showed that a dimensionless effect size, closely related to our discriminability ratio, was a much better predictor of power than the number of taxa.
Moreover, they found that when discriminability ratio is small, power can not be greatly improved by increasing the number of taxa sampled, even to very large numbers.
This finding echoes our own results (Figs.~\ref{fig:eta-phi-power-contours} and \ref{fig:snr-power-contours}).

\subsection*{Reliability of parameter estimates}

Our results show that the parameters differed in the accuracy with which they could be estimated.
Relatively accurate estimates of $\phi$ were common when trees were of moderate size and $\phi$ was not very small (Fig.~\ref{fig:eta-phi-phi-contours}).
Selection opportunity, $\eta$, was the most difficult to estimate (Fig.~\ref{fig:eta-phi-eta-contours}):
relative RMSE was less than 50\% only when both selection opportunity and discriminability ratio were high.

It is noteworthy, however, that relative error of 50\% and more were common in many regions of parameter space.
Thus, our results call for caution in the interpretation of parameter estimates for OUCH-class models.
Moreover, since as our results show, the quality of parameter estimates varies markedly across parameter space, a case-by-case approach to the quantification of uncertainty in parameter estimates is essential.
For this purpose, approaches based on the parametric bootstrap \citep{Scales:2009,Boettiger:2012} are recommended.

Estimates of the trait optima tended to be more accurate than estimates of $\eta$ and $\phi$, reflecting the fact that the three optima must be distinguishable for the OU3 model to be well supported.
We noted that the identity of the regime operating at the root influenced the estimates of the selective optima (Fig.~\ref{fig:theta-RMSE-by-root}).
In particular, the best-estimated regime was the regime operative at the root and the worst-estimated regime was that with the optimum farthest from that of the root regime.

\citet{Ho2013} demonstrated mathematically, for the single-optimum OU process on an ultrametric tree, that the parameters $\alpha$ and $\sigma^2$ can be consistently estimated, but that the estimate of $\sigma^2$ converges faster than does the estimate of $\alpha$.
[An estimator is said to be consistent when estimates converge to the true value as the number of data increases.]
This parallels our result that, though the estimates of both $\eta$ and $\phi$ improve with tree size, $\phi$ is always better estimated (Figs.~\ref{fig:eta-phi-phi-contours} and \ref{fig:eta-phi-eta-contours}).

In later work, \citet{Ho2014} proved that selective optima can be consistently estimated as long as, for some selective regime, the branches in that regime do not form a connected subtree.
Our finding that the accuracy of trait-optima estimates increases with the size of the tree (Fig.~\ref{fig:eta-phi-thetaC-contours}) is consistent with the results of \citet{Ho2014}, since all trees considered in our study met that criterion.
We expect that accurate estimates of the trait optima would be impossible otherwise.

\subsection*{Parameter estimation and the topography of the likelihood surface}

\begin{figure}
\includegraphics[width=\linewidth]{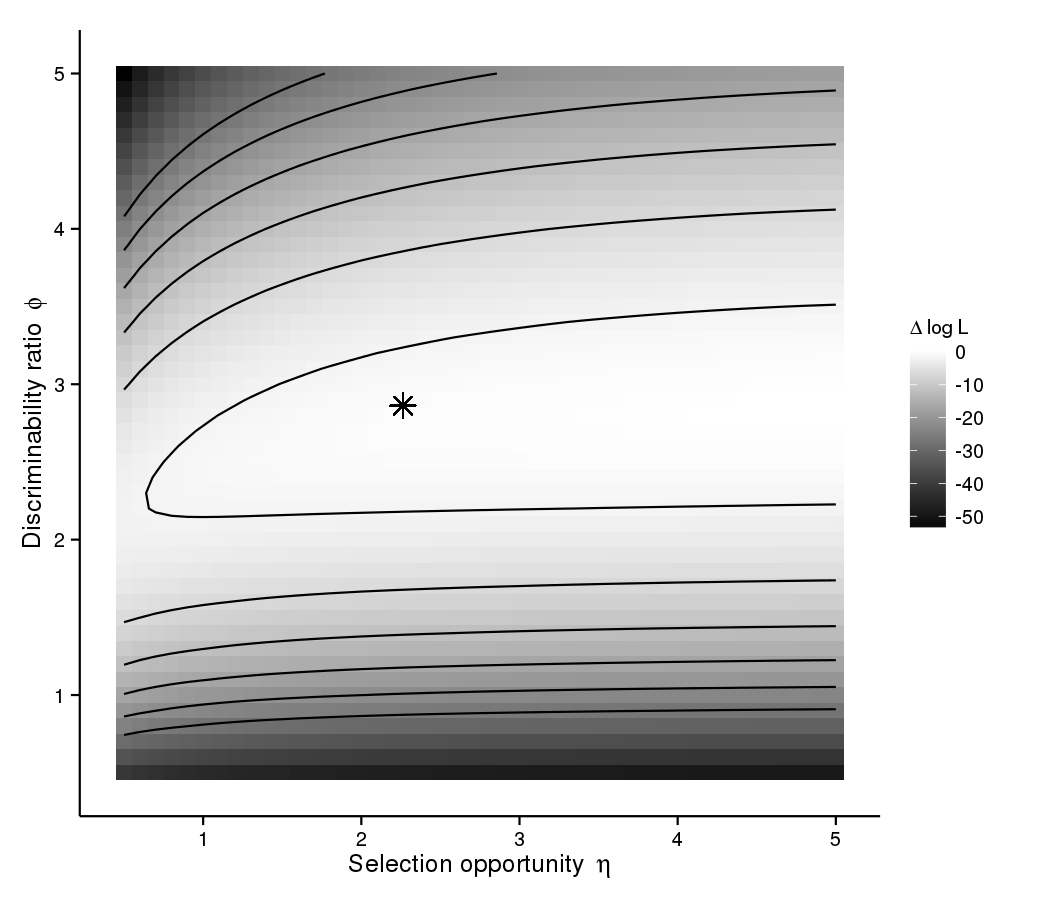} \hfill{}
\caption{
  Typical contour plot of log likelihood against $\eta$ and $\phi$.
  Greyscale and contours show log likelihood relative to its maximum value.
  For each $\eta$ and $\phi$ combination, the log likelihood was calculated for the OU3 model without fitting the model to the data.
  The data generated were for a tree of size 39, with true $\eta$ and $\phi$ shown by the asterisk.
}
\label{fig:likridge}
\end{figure}

It is perhaps striking that, in certain regions of parameter space, model selection can be quite reliable, yet parameter estimates fraught with error (cf.~Figs.~\ref{fig:eta-phi-power-contours}--\ref{fig:eta-phi-eta-contours}).
This is due to inherent limits in the identifiability of the parameters, i.e., to limits in the information content of the data.
Fig.~\ref{fig:likridge} shows the shape of the likelihood surface in one typical case.
The relatively high curvature in the $\phi$ direction corresponds to the fact that estimates of that parameter can be made with some precision.
The likelihood ridge extending in the $\eta$ direction implies that many values of $\eta$ are roughly equally consistent with the data:
selection opportunity is much less well identified.
The existence of likelihood ridges has been noted since Hansen's original (\citeyear{Hansen:1997}) paper, but is still not widely appreciated \citep{Ho2014}.
However, since model selection depends on the maximum height attained by the surface, and not on its shape, discrimination between models can be sharp even when parameter estimates are not.
Such ridges in likelihood space lead to parameter estimates with large confidence intervals and caution against over-interpretation of parameter estimates.
It is in general worthwhile to explore the likelihood surface for a given dataset by evaluating the likelihood at parameter values near the MLE.

The foregoing observations have implications for more parameter-rich methods based on generalizations of the OUCH class in which parameters are allowed to vary across the tree \citep{OMeara:2006,Beaulieu:2012}.
Our study shows that the information contained in the data are already markedly limited when these parameters are constant.
Increasing the number of parameters can only further dilute information, leading to even greater uncertainty in estimates.

\subsection*{Assessing reliability in practice}

\begin{figure}
\includegraphics[width=\linewidth]{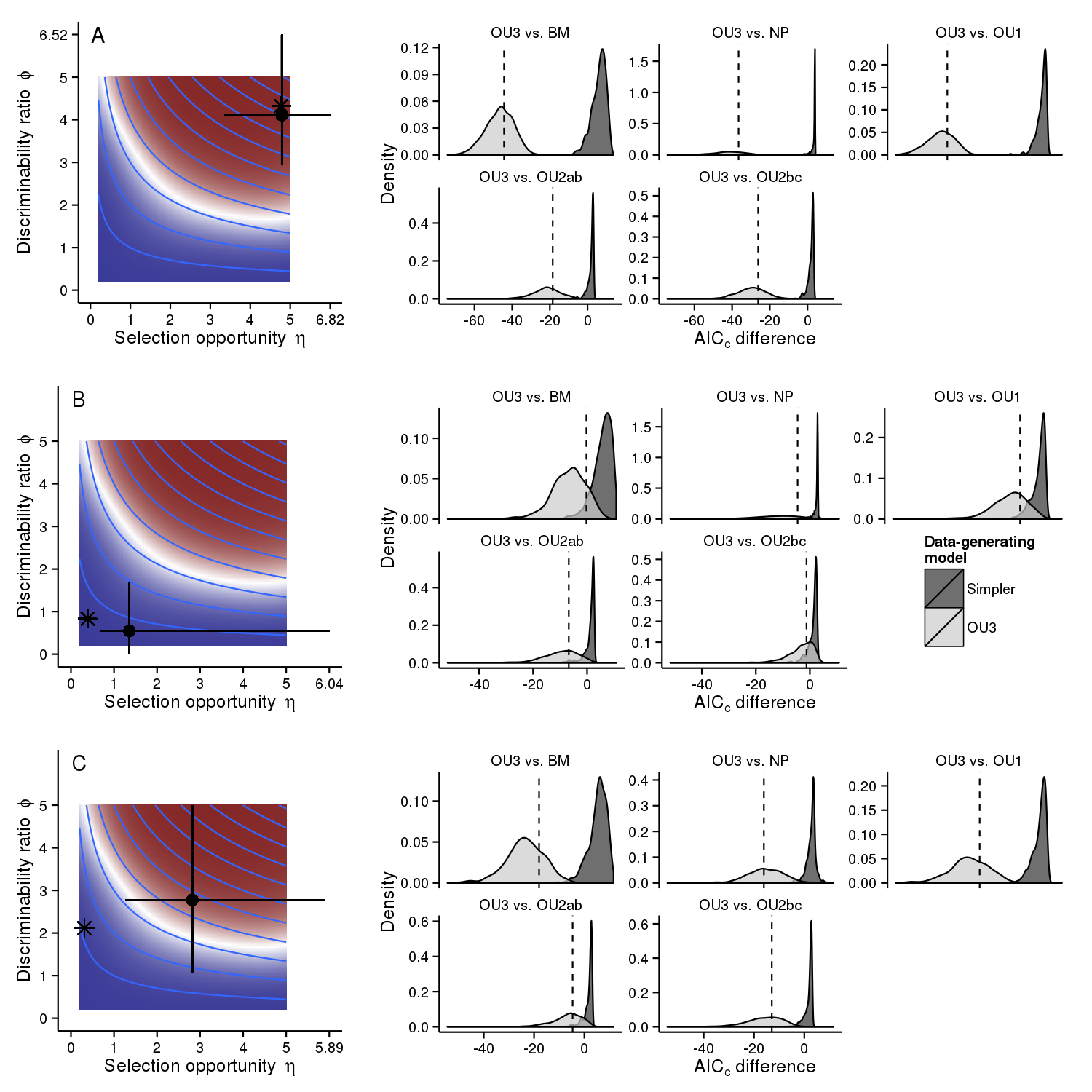} \hfill{}
\caption{
  \small
  Three scenarios illustrating various parameter estimation and model selection outcomes, as described in the text.
  Scenario A is drawn from a region where power is high and parameter estimates reliable. Scenarios B and C are drawn from regions where power is low and parameter estimation unreliable.
  In Scenario B, both the true parameters and the estimates lie in regions of low power.
  In Scenario C, however, the overestimation of $\eta$ and $\phi$ tempts one to the specious conclusion that the true parameters are in a region of high power and accuracy.
  The plots at left indicate true parameter values (asterisks) and ML estimates (filled circles).
  The crosshairs shows 95\% bootstrap confidence intervals.
  The plots at right show the phylogenetic Monte Carlo results for pairwise comparison of alternative models based on 1000 bootstrap replicates.
  The density plots show the distribution of $\mathrm{AIC}_c$ differences between the indicated models (OU3-simpler model).
  In each case, the light gray density shows the distribution of $\mathrm{AIC}_c$ differences when the data-generating process is OU3 at its MLE; the dark gray differences result when the data-generating process is the simpler model at its MLE.
  The vertical dashed line shows the observed $\mathrm{AIC}_c$ difference on the original data.
  Comparing the observed $\mathrm{AIC}_c$ difference to the dark gray density gives an indication of the statistical significance of the observed $\mathrm{AIC}_c$ difference.
  Lack of overlap between the light gray and dark gray densities indicates high power of the hypothesis test.
}
\label{fig:pmc}
\end{figure}

While our simulation study sheds some light on how phylogenetic comparative analyses based on OUCH models can be expected to perform in principle, some care must be taken in translating these results into practical recommendations.
For example, having fit an OUCH model to one's data, one might be tempted to map one's estimates $\hat{\eta}$, $\hat{\phi}$ onto Fig.~\ref{fig:eta-phi-power-contours} in an attempt to gauge the reliability of one's model selection result.
Similarly, one might be tempted to assume that unambiguously strong support for an adaptive model implies reliability of parameter estimates.
Our results suggest that these temptations should be resisted, as inaccuracy in parameter estimates can lead to misleading conclusions.
Moreover, such inaccuracy may not be readily detectable.
We illustrate the variety of potential outcomes by focusing on three scenarios (Fig.~\ref{fig:pmc}).


In the optimistic Scenario A, $\hat{\eta}$ and $\hat{\phi}$ are both large and the true values of these parameters are well within the 95\% bootstrap confidence interval.
The phylogenetic Monte Carlo distributions \citep{Scales:2009,Boettiger:2012} indicate that the $\mathrm{AIC}_c$-based rejection of the simpler alternatives is justifiable at a high confidence level.
This scenario accords with our expectation (based on Figs.~\ref{fig:eta-phi-power-contours}--\ref{fig:eta-phi-eta-contours} and \ref{fig:snr-power-contours}):
when the true values of $\eta$ and $\phi$ are both large, model misidentification is rare and parameter estimates tend to be accurate.

Scenario B is likewise optimistic.
Here, though according to our $\mathrm{AIC}_c$ criterion the OU3 model is best supported, $\hat{\eta}$ and $\hat{\phi}$ are both low.
In accordance with the simulation study results, bootstrap confidence intervals are large.
The simulation study also suggests, however, that in this region of parameter space, selection of the OU3 model should be rare and unreliable.
This is borne out by the phylogenetic Monte Carlo analysis, which shows that the observed $\mathrm{AIC}_c$ difference frequently occurs even when the data are generated by the simpler models.
Thus, in this scenario, both the parameter estimates and the phylogenetic Monte Carlo results warn us against placing too much confidence in our selection of the OU3 model.

Scenario C yields a cautionary tale.
As in scenario B, the data were generated in a regime where, according to the simulation study, reliable recovery of the data-generating model should be rare.
For the particular data observed in this scenario, however, OU3 was indeed best supported, with parameters estimated as shown.
These parameter estimates are far from the truth, incorrectly suggesting that we are within the high-power, high-accuracy regime.
Nor does phylogenetic Monte Carlo (Fig.~\ref{fig:pmc}C) give us any hint of our mistake.
Scenario C was rare in our simulation study (for example, fewer than 10\% of datasets with true $\eta$ and $\phi$ less than 2 resulted in estimates $\hat{\eta}$ and $\hat{\phi}$ larger than 2 and bias greater than 10\%).
Its existence, however, urges a degree of caution in the interpretation of parameter estimates in this model and generalizations of it that are even more parameter-rich.
This scenario illustrates that while parameter estimation and model selection are intimately related, it is possible to have wildly inaccurate parameter estimates while model selection works well.

\subsection*{The utility and limitations of the dimensionless model}

Our analysis shows that the model selection power and parameter estimation accuracy are largely determined by the dimensionless parameters $\eta$ and $\phi$ and by the number of taxa sampled.
Because these quantities are dimensionless, they can be meaningfully interpreted and compared across studies, whereas the interpretation and comparison of parameters like $\alpha$ and $\sigma$ are complicated by their dependence on the measurement scales of time and trait \citep[see also ][]{Ho2014}.

Although the formulation of the models in terms of dimensionless parameters is useful, it does not translate directly into a prescription for data analysis.
In particular, we note that, although $\eta=\alpha\,T$ is unambiguous if the tree depth, $T$, is known, the discriminability ratio, $\phi$, depends on the unknown quantity $\Delta\theta$.
Moreover, as we saw above, the potential for inaccurate parameter estimates means that, even having estimated the optima (and hence $\Delta\theta$), one cannot safely conclude that one is well within a region where estimates are reliable and power is high.
In effect, the simulation study described here maps out an important portion of the model parameter space.
In contemplating the use of OUCH-class methods for data analysis, we do well to understand this map, even if we must live with the fact that we cannot always know with certainty where we are on it.

\subsection*{Beyond OUCH}

When OUCH models are used for the purpose of hypothesis testing, a difficulty often encountered is that the biological hypothesis one wishes to test cannot be uniquely translated into a single regime painting.
For example, it is often the case that the history of selective regimes deep in the phylogeny is not known and various regime paintings are therefore plausible.
There is no difficulty when a hypothesis does entail a unique regime painting: it is a straightforward matter to obtain maximum-likelihood (ML) parameters and a maximized likelihood that can be used to compare the hypothesis against alternatives.
Should this likelihood be sufficiently low, one may reject the original hypothesis, precisely because the regime painting is its necessary consequence.
However, if the hypothesis is compatible with more than one regime painting, rejection of any particular painting does not logically force rejection of the hypothesis.
In other words, in such a situation, uncertainty concerning the regime painting remains.
Many recent methodological developments have focused on accounting for this uncertainty in the regime painting \citep{Revell:2011,Eastman:2011,Hipp:2010,Ingram:2013,Uyeda2014}.
It is worth considering here how such uncertainty can properly be taken into account.
The two major branches of statistical philosophy offer two closely related approaches.

A Bayesian approach would be to treat the regime painting, $P$, as a additional parameter, assign priors, and perform inference as usual \citep{Bollback:2006}.
The key ingredient in doing so would be the proportionality relation
\begin{equation}\label{eq:bayes1}
  \pi(\Theta,P|X) \propto f(X|\Theta,P)\,\pi(\Theta,P),
\end{equation}
where $X$ is the data, $\Theta$ stands for the parameters of the OU (or BM) phenotype model, $\pi(\Theta,P|X)$ is the posterior distribution, $\pi(\Theta,P)$ is the joint prior on $P$ and $\Theta$, and $f(X|\Theta,P)$ is the OU (or BM) likelihood.
One drawback of this approach is the need to specify priors on $\Theta$ and $P$ even when little or no prior information on these parameters exists.
The methods proposed by \citet{Revell:2011}, \citet{Eastman:2011}, and \citet{Uyeda2014} adopt this approach.
If one is willing to propose a probabilistic model for $P$, however, one arrives at a hierarchical model, for which the key proportionality relation is
\begin{equation}
\pi(\Theta,\Phi,P|X) \propto f(X|\Theta,P)\,g(P|\Phi)\,\pi(\Theta,\Phi),
\end{equation}
where $g(P|\Phi)$ is the likelihood of regime painting $P$ under the parameters $\Phi$.
One still has the difficulty of assigning priors over $\Theta$ and $\Phi$, but this may be somewhat more straightforward than doing so over the regime paintings $P$.

The ML approach is to integrate over the uncertainty in $P$.
Specifically, one seeks parameters $\Theta$ and $\Phi$ that maximize the likelihood
\begin{equation}\label{eq:ml1}
\ell(X|\Theta,\Phi) = \sum f(X|\Theta,P)\,g(P|\Phi),
\end{equation}
the sum being taken over all regime paintings $P$.
Because the space of paintings will often be large, a Monte Carlo approach to the approximation of the sum may often be needed in practice.
As with the Bayesian approach, accurate estimation requires adequate exploration of the space of $P$.
Though the ML approach requires that one must maximize the likelihood (Eq.~\ref{eq:ml1}), which can sometimes be difficult relative to the problem of sampling from the posterior, it has the advantage that no priors are needed.

In other words, the maximum likelihood framework we have applied in this paper can be extended to deal with regime-painting uncertainty.
One must translate one's hypothesis into a probability distribution over regime paintings.
Then, to compute the likelihood at any given set of parameters, one proceeds as follows.
First, generate a large number of regime paintings from the regime-painting model.
For each painting, $P$, compute the conditional likelihood of the data under the phenotype model (e.g., of OUCH type), given $P$.
The average of these conditional likelihoods is the required likelihood.
This recipe gives the likelihood at a given point in parameter space:
to complete the inference, it remains to maximize it over all values of the parameters.

While certain theoretical advantages attach to the Bayesian or ML prescriptions (both of them based on likelihood), other approaches are of course possible.
One might, for example, propose a method that seeks the ``true'' regime painting by searching over the space of all such paintings in quest of the one that yields the highest maximized likelihood, lowest AIC, or is by some other criterion deemed best \citep{Ingram:2013}.
One worries that, since the size of the parameter space is large, such a method would suffer from overfitting and statistical inconsistency of estimates.
Another alternative might be to propose a large number of paintings, maximize the parameters $\Theta$ over each one, and somehow average the resulting parameters and maximized likelihoods \citep{Collar2009a,Collar2010b,Hipp:2010}.
Note that this is not at all the same as the integration over the uncertainty in $P$ performed under either of the likelihood-based prescriptions: the maximization and averaging steps are transposed.

Clearly, there is growing interest in the development and use of sophisticated phylogenetic comparative methods based on models more complex and realistic than the OUCH-class models studied here.
While our results inform some expectations, a thorough study of the statistical properties of these methods would be invaluable.
Overall, we suggest that it is imperative for biologists to understand not only the potential of these methods, but also their inherent limitations.

\section*{Acknowledgements}

Our sincere thanks to the editor, associate editor, and three anonymous reviewers, whose critical feedback greatly improved this manuscript.
We also thank Julio Rivera, Jeffrey Scales, Yvonne Chan, Elizabeth Henry, Bob Thomson, and members of the Fall 2012 Phylogenetic Discussion Group at the University of Hawai'i for valuable comments on early versions of this work.
This research was supported by NSF awards DEB~1145733 and DEB~0515390 (to MAB) and DEB~0542360 (to AAK).
This paper is University of Hawai'i Department of Biology contribution number 20XX-XX.

\bibliographystyle{sysbio}
\bibliography{ouchsim}

\begin{thebibliography}{53}
\providecommand{\natexlab}[1]{#1}
\providecommand{\selectlanguage}[1]{\relax}

\bibitem[{An\'{e}(2008)}]{Ane:2008}
An\'{e}, C. 2008. Analysis of comparative data with hierarchical
  autocorrelation. Annals of Applied Statistics 2:1078--1102.

\bibitem[{Arnold and Wade(1984)}]{Arnold1984}
Arnold, S.~J. and M.~J. Wade. 1984. On the measurement of natural and sexual
  selection: Applications. Evolution 38:720--734.

\bibitem[{Bartoszek et~al.(2012)Bartoszek, Pienaar, Mostad, Andersson, and
  Hansen}]{Bartoszek:2012}
Bartoszek, K., J.~Pienaar, P.~Mostad, S.~Andersson, and T.~F. Hansen. 2012. A
  phylogenetic comparative method for studying multivariate adaptation. Journal
  of Theoretical Biology 314:204--215.

\bibitem[{Beaulieu et~al.(2012)Beaulieu, Jhwueng, Boettiger, and
  O'Meara}]{Beaulieu:2012}
Beaulieu, J.~M., D.-C. Jhwueng, C.~Boettiger, and B.~C. O'Meara. 2012. Modeling
  stabilizing selection: expanding the ornstein-uhlenbeck model of adaptive
  evolution. Evolution 66:2369--2383.

\bibitem[{Boettiger et~al.(2012)Boettiger, Coop, and Ralph}]{Boettiger:2012}
Boettiger, C., G.~Coop, and P.~Ralph. 2012. Is your phylogeny informative?
  measuring the power of comparative methods. Evolution 66:2240--2251.

\bibitem[{Bollback(2006)}]{Bollback:2006}
Bollback, J.~P. 2006. Simmap: Stochastic character mapping of discrete traits
  on phylogenies. BMC Bioinformatics 7:88.

\bibitem[{Buckingham(1914)}]{Buckingham1914}
Buckingham, E. 1914. On physically similar systems; illustrations of the use of
  dimensional equations. Physical Review 4:345--376.

\bibitem[{Burnham and Anderson(2002)}]{Burnham:2002}
Burnham, K.~P. and D.~R. Anderson. 2002. Model selection and multimodel
  inference: A practical information-theoretic approach. Second ed.,
  Springer-Verlag.

\bibitem[{Butler and King(2004)}]{Butler:2004}
Butler, M.~A. and A.~A. King. 2004. Phylogenetic comparative analysis: a
  modeling approach for adaptive evolution. American Naturalist 164:683--695.

\bibitem[{Collar et~al.(2009)Collar, O'Meara, Wainwright, and
  Near}]{Collar2009a}
Collar, D.~C., B.~C. O'Meara, P.~C. Wainwright, and T.~J. Near. 2009. Piscivory
  limits diversification of feeding morphology in centrarchid fishes. Evolution
  63:1557--1573.

\bibitem[{Collar et~al.(2011)Collar, Schulte~II, and Losos}]{Collar2011}
Collar, D.~C., J.~A. Schulte~II, and J.~B. Losos. 2011. Evolution of extreme
  body size disparity in monitor lizards ({V}aranus). Evolution 65:2664--2680.

\bibitem[{Collar et~al.(2010)Collar, Schulte~II, O'Meara, and
  Losos}]{Collar2010b}
Collar, D.~C., J.~A. Schulte~II, B.~C. O'Meara, and J.~B. Losos. 2010. Habitat
  use affects morphological diversification in dragon lizards. Journal of
  Evolutionary Biology 23:1033--1049.

\bibitem[{Eastman et~al.(2011)Eastman, Alfaro, Joyce, Hipp, and
  Harmon}]{Eastman:2011}
Eastman, J.~M., M.~E. Alfaro, P.~Joyce, A.~L. Hipp, and L.~J. Harmon. 2011. A
  novel comparative method for identifying shifts in the rate of character
  evolution on trees. Evolution 65:3578--89.

\bibitem[{Edwards and Smith(2010)}]{Edwards2010}
Edwards, E.~J. and S.~A. Smith. 2010. Phylogenetic analyses reveal the shady
  history of c4 grasses. Proceedings of the National Academy of Sciences of the
  U.S.A. 107:2532--2537.

\bibitem[{Felsenstein(1973)}]{Felsenstein:1973}
Felsenstein, J. 1973. Maximum-likelihood estimation of evolutionary trees from
  continuous characters. American Journal of Human Genetics 25:471--492.

\bibitem[{Felsenstein(1985)}]{Felsenstein:1985}
Felsenstein, J. 1985. Phylogenies and comparative method. American Naturalist
  125:1--15.

\bibitem[{Gomez and Th\'{e}ry(2007)}]{Gomez2007}
Gomez, D. and M.~Th\'{e}ry. 2007. Simultaneous crypsis and conspicuousness in
  color patterns: comparative analysis of a neotropical rainforest bird
  community. American Naturalist 169:S42----S61.

\bibitem[{Hansen(1997)}]{Hansen:1997}
Hansen, T.~F. 1997. Stabilizing selection and the comparative analysis of
  adaptation. Evolution 51:1341--1351.

\bibitem[{Hansen et~al.(2008)Hansen, Pienaar, and Orzack}]{Hansen:2008}
Hansen, T.~F., J.~Pienaar, and S.~H. Orzack. 2008. A comparative method for
  studying adaptation to a randomly evolving environment. Evolution
  62:1965--1977.

\bibitem[{Harmon et~al.(2008{\natexlab{a}})Harmon, Weir, Brock, Glor, and
  W.}]{Harmon:2008}
Harmon, L.~J., J.~Weir, C.~Brock, R.~E. Glor, and C.~W. 2008{\natexlab{a}}.
  Geiger: Investigating evolutionary radiations. Bioinformatics (Oxford,
  England) 24:129--131.

\bibitem[{Harmon et~al.(2008{\natexlab{b}})Harmon, Weir, Brock, Glor, and
  Challenger}]{Harmon2008}
Harmon, L.~J., J.~T. Weir, C.~D. Brock, R.~E. Glor, and W.~Challenger.
  2008{\natexlab{b}}. {GEIGER: Investigating evolutionary radiations}.
  Bioinformatics 24:129--131.

\bibitem[{Hipp and Escudero(2010)}]{Hipp:2010}
Hipp, A.~L. and M.~Escudero. 2010. Maticce: mapping transitions in continuous
  character evolution. Bioinformatics (Oxford, England) 26:132--133.

\bibitem[{Ho and An\'{e}(2013)}]{Ho2013}
Ho, L. S.~T. and C.~An\'{e}. 2013. Asymptotic theory with hierarchical
  autocorrelation: Ornstein-uhlenbeck tree models. Annals of Statistics
  41:957--981.

\bibitem[{Ho and An\'{e}(2014)}]{Ho2014}
Ho, L. S.~T. and C.~An\'{e}. 2014. Intrinsic inference difficulties for trait
  evolution with {O}rnstein-{U}hlenbeck models. Methods Ecol. Evol.
  2:1133--1146.

\bibitem[{Hunt(2006)}]{Hunt2006}
Hunt, G. 2006. {Fitting and comparing models of phyletic evolution: random
  walks and beyond}. Paleobiology 32:578--601.

\bibitem[{Ingram and Mahler(2013)}]{Ingram:2013}
Ingram, T. and D.~L. Mahler. 2013. Surface: detecting convergent evolution from
  comparative data by fitting ornstein-uhlenbeck models with stepwise akaike
  information criterion. Methods in Ecology and Evolution page in press.

\bibitem[{King(2008)}]{subplex}
King, A.~A. 2008. subplex: Subplex optimization algorithm. R package version
  1.1-3.

\bibitem[{King and Butler(2009)}]{King:2009}
King, A.~A. and M.~A. Butler. 2009. ouch: Ornstein-Uhlenbeck models for
  phylogenetic comparative hypotheses.

\bibitem[{Kirkpatrick and Slatkin(1993)}]{Kirkpatrick1993}
Kirkpatrick, M. and M.~Slatkin. 1993. Searching for evolutionary patterns in
  the shape of a phylogenetic tree. Evolution 47:1171--1181.

\bibitem[{Kozak and Wiens(2010)}]{Kozak2010a}
Kozak, K.~H. and J.~J. Wiens. 2010. Niche conservatism drives elevational
  diversity patterns in {A}ppalachian salamanders. American Naturalist
  176:40--54.

\bibitem[{Labra et~al.(2009)Labra, Pienaar, and Hansen}]{Labra2009}
Labra, A., J.~Pienaar, and T.~F. Hansen. 2009. Evolution of thermal physiology
  in \emph{Liolaemus} lizards: adaptation, phylogenetic inertia, and niche
  tracking. American Naturalist 174:204--220.

\bibitem[{Lande(1976)}]{Lande:1976}
Lande, R. 1976. Natural selection and random genetic drift in phenotypic
  evolution. Evolution 30:314--334.

\bibitem[{Lande(1980)}]{Lande:1980}
Lande, R. 1980. Genetic variation and phenotypic evolution during allopatric
  speciation. American Naturalist 114:463--479.

\bibitem[{Monteiro and Nogueira(2011)}]{Monteiro2011}
Monteiro, L.~R. and M.~R. Nogueira. 2011. Evolutionary patterns and processes
  in the radiation of phyllostomid bats. BMC Evolutionary Biology 11:137.

\bibitem[{Mooers and Heard(1997)}]{Mooers1997}
Mooers, A.~O. and S.~B. Heard. 1997. Inferring evolutionary process from
  phylogenetic tree shape. Quarterly Review of Biology 72:31--54.

\bibitem[{Murray(2002)}]{Murray2002}
Murray, J.~D. 2002. Mathematical Biology. {I}: An Introduction, vol.~17 of
  \emph{Interdisciplinary Applied Mathematics}. 3rd ed., Springer-Verlag,
  Berlin.

\bibitem[{Nicholson et~al.(2005)Nicholson, Glor, Kolbe, Larson, Hedges, and
  Losos}]{Nicholson:2005kn}
Nicholson, K.~E., R.~E. Glor, J.~J. Kolbe, A.~Larson, S.~B. Hedges, and J.~B.
  Losos. 2005. Mainland colonization by island lizards. Journal of biogeography
  32:929--938.

\bibitem[{O'Meara et~al.(2006)O'Meara, An\'{e}, Sanderson, and
  Wainwright}]{OMeara:2006}
O'Meara, B.~C., C.~An\'{e}, M.~J. Sanderson, and P.~C. Wainwright. 2006.
  Testing for different rates of continuous trait evolution using likelihood.
  Evolution 60:922--933.

\bibitem[{Paradis et~al.(2004)Paradis, Claude, and Strimmer}]{Paradis:2004}
Paradis, E., J.~Claude, and K.~Strimmer. 2004. Ape: Analyses of phylogenetics
  and evolution in r language. Bioinformatics (Oxford, England) 20:289--290.

\bibitem[{Press et~al.(1992)Press, Teukolsky, Vetterling, and
  Flannery}]{Press1992}
Press, W.~H., S.~A. Teukolsky, W.~T. Vetterling, and B.~P. Flannery. 1992.
  Numerical Recipes in C. Cambridge University Press.

\bibitem[{{R Core Team}(2013)}]{R}
{R Core Team}. 2013. R: A Language and Environment for Statistical Computing. R
  Foundation for Statistical Computing, Vienna, Austria.

\bibitem[{Revell et~al.(2011)Revell, Mahler, Peres-Neto, and
  Redelings}]{Revell:2011}
Revell, L.~J., D.~L. Mahler, P.~R. Peres-Neto, and B.~D. Redelings. 2011. A new
  phylogenetic method for identifying exceptional phenotypic diversification.
  Evolution 66:135--146.

\bibitem[{Rowan(1990)}]{Rowan:1990}
Rowan, T. 1990. Functional Stability Analysis of Numerical Algorithms. Ph.D.
  thesis, Department of Computer Sciences, University of Texas at Austin.

\bibitem[{Sackin(1972)}]{Sackin1972}
Sackin, M.~J. 1972. {``Good''} and ``bad'' phenograms. Systematic Zoology
  21:225--226.

\bibitem[{Scales et~al.(2009)Scales, King, and Butler}]{Scales:2009}
Scales, J.~A., A.~A. King, and M.~A. Butler. 2009. Running for your life or
  running for your dinner: what drives fiber-type evolution in lizard locomotor
  muscles? American Naturalist 173:543--553.

\bibitem[{Setiadi et~al.(2011)Setiadi, McGuire, Brown, Zubairi, Iskandar,
  Andayani, Supriatna, and Evans}]{Setiadi2011b}
Setiadi, M.~I., J.~a. McGuire, R.~M. Brown, M.~Zubairi, D.~T. Iskandar,
  N.~Andayani, J.~Supriatna, and B.~J. Evans. 2011. Adaptive radiation and
  ecological opportunity in sulawesi and philippine fanged frog (limnonectes)
  communities. American Naturalist 178:221--240.

\bibitem[{Simpson(1953)}]{Simpson:1953}
Simpson, G.~G. 1953. The major features of evolution, vol. no. 17. Columbia
  University Press, New York.

\bibitem[{Strang(1986)}]{Strang1986}
Strang, G. 1986. Introduction to Applied Mathematics. Wellesley-Cambridge
  Press, Wellesley, Mass.

\bibitem[{Uyeda and Harmon(2014)}]{Uyeda2014}
Uyeda, J.~C. and L.~J. Harmon. 2014. A novel bayesian method for inferring and
  interpreting the dynamics of adaptive landscapes from phylogenetic
  comparative data. Syst Biol 63:902--918.

\bibitem[{Venditti et~al.(2011)Venditti, Meade, and Pagel}]{Venditti:2011}
Venditti, C., A.~Meade, and M.~Pagel. 2011. Multiple routes to mammalian
  diversity. Nature 479:393--396.

\bibitem[{Whittall and Hodges(2007)}]{Whittall2007}
Whittall, J.~B. and S.~a. Hodges. 2007. Pollinator shifts drive increasingly
  long nectar spurs in columbine flowers. Nature 447:706--709.

\bibitem[{Wood(2003)}]{Wood2003}
Wood, S.~N. 2003. Thin plate regression splines. Journal of the Royal
  Statistical Soceity B 65:95--114.

\bibitem[{Wood(2006)}]{Wood2006}
Wood, S.~N. 2006. Generalized Additive Models: An Introduction with R. CRC
  Press.

\end{thebibliography}

\clearpage

\appendix

\renewcommand{\thesection}{\Alph{section}} \setcounter{section}{0}
\setcounter{figure}{0}
\renewcommand{\thefigure}{A\arabic{figure}}

\section{Supplementary figures}
\label{sec:supplementaryplots}

\begin{figure}[b]
  \includegraphics[width=\textwidth]{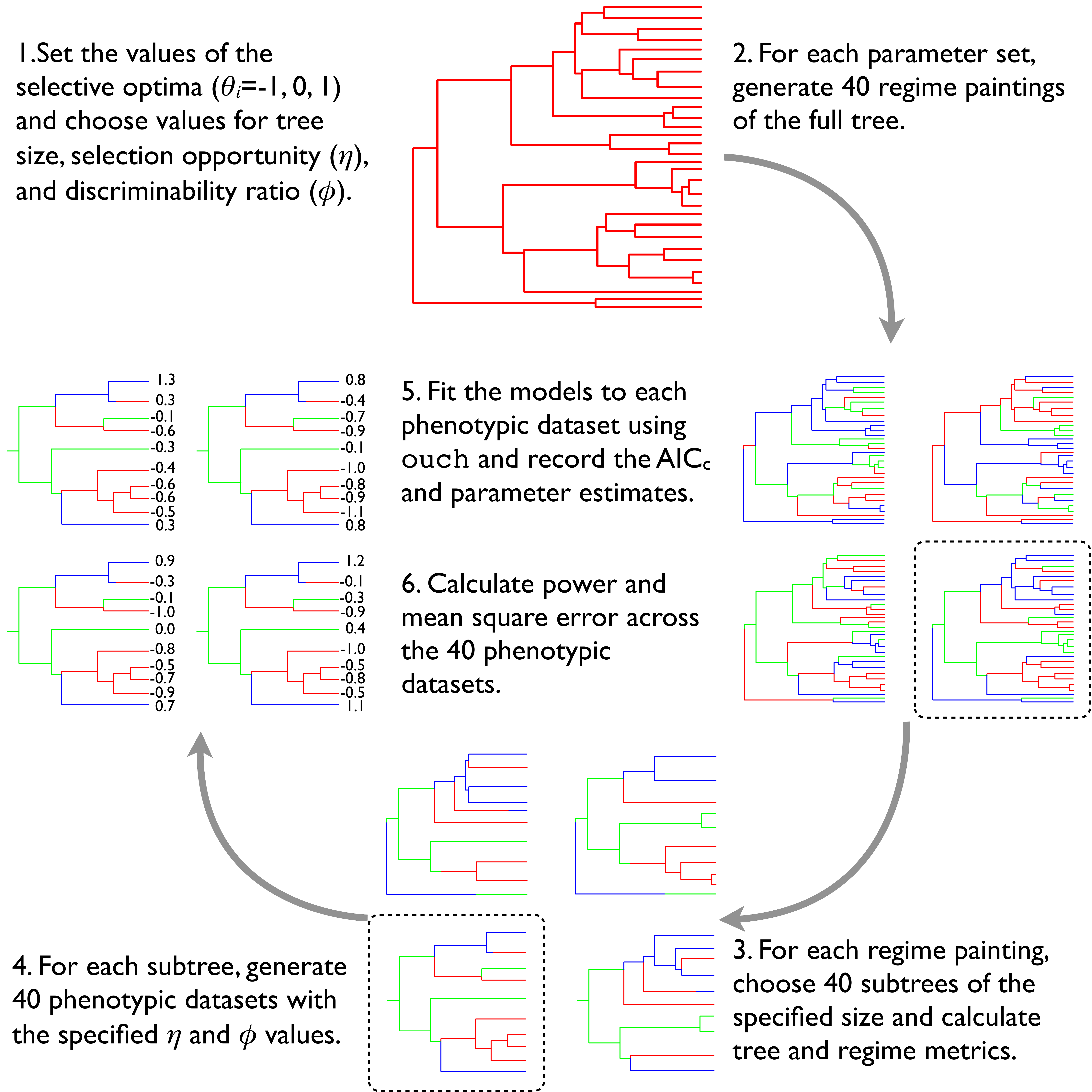}
  \caption{Plan of the simulation study.}
  \label{fig:outline}
\end{figure}

\begin{figure}
\includegraphics[width=\linewidth]{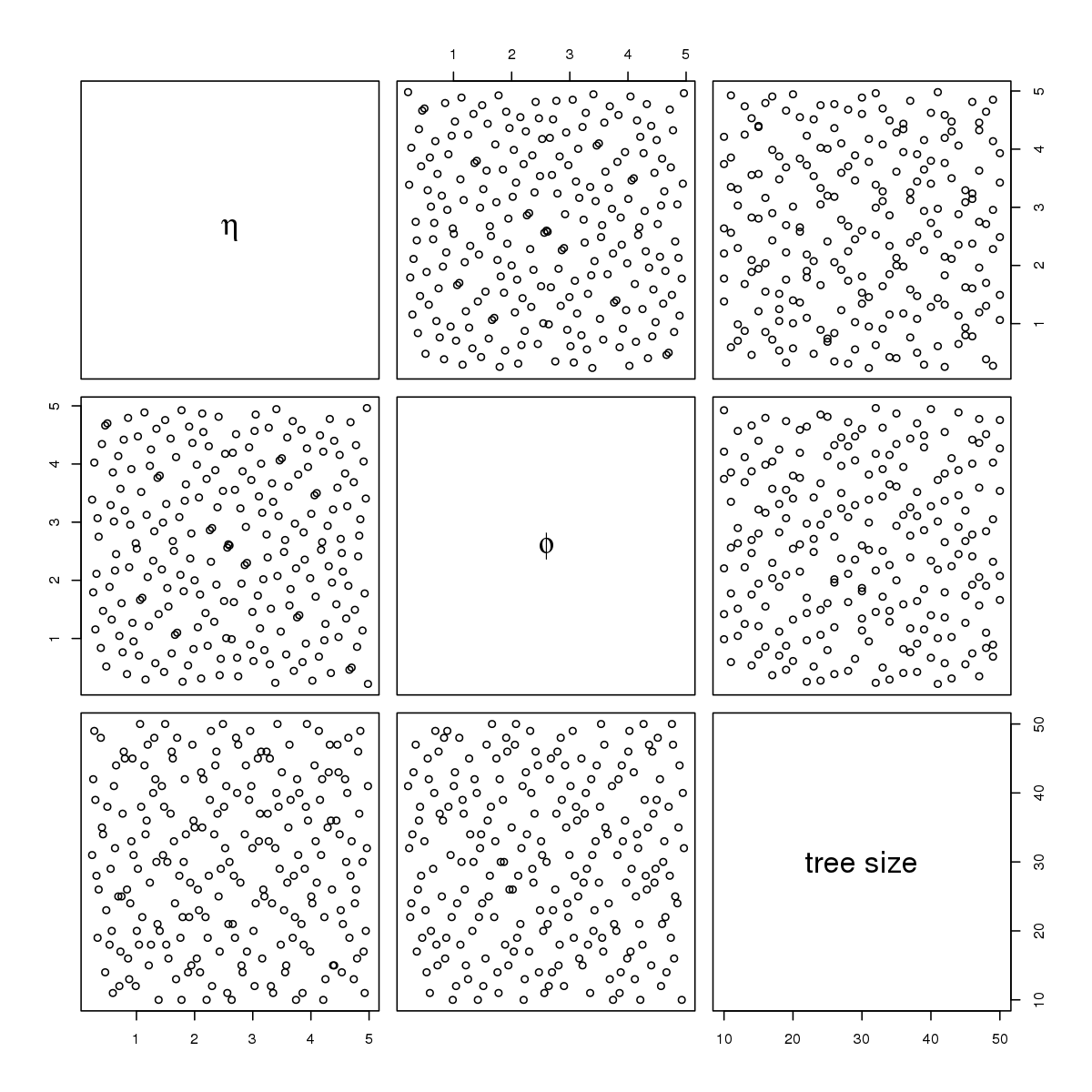} \hfill{}
\caption{Scatterplots showing the distribution of selection opportunity $\eta$, discriminability ratio $\phi$, and tree size values explored in this study.}
\label{fig:paramcombs}
\end{figure}

Fig.~\ref{fig:outline} is a schematic diagram of the simulation study design.
Fig.~\ref{fig:paramcombs} shows the distribution of selection opportunity $\eta$, discriminability ratio $\phi$, and tree size explored in this simulation study.
As explained in the main text, we used a Sobol' low-discrepancy sequence \citep{Press1992} to generate 220 $(\eta,\phi,\text{tree size})$ triples.
This samples in such a way that each variable is sampled uniformly across its range and such that combinations of variables are also sampled uniformly.
This range of selection opportunity $\eta$ values corresponds to values of $\alpha$ ranging between 0.2 and 5 (because $\eta=\alpha~T$).
Correspondingly, this implies that the depth of the phylogeny, ranges between 0.29 and 7.2 phylogenetic half-lives \citep{Hansen:1997}.
This range of discriminability ratio $\phi$ values corresponds to a range of noise intensity $\sigma$ values between 0.13 and 16 (because $\phi = \frac{\sqrt{2~\alpha}\Delta\theta}{\sigma}$).
Correspondingly, this implies that the separation between optima varies between 0.2 and 5 standard deviations of the stationary distribution around each optimum.

\begin{figure}
\includegraphics[width=\linewidth]{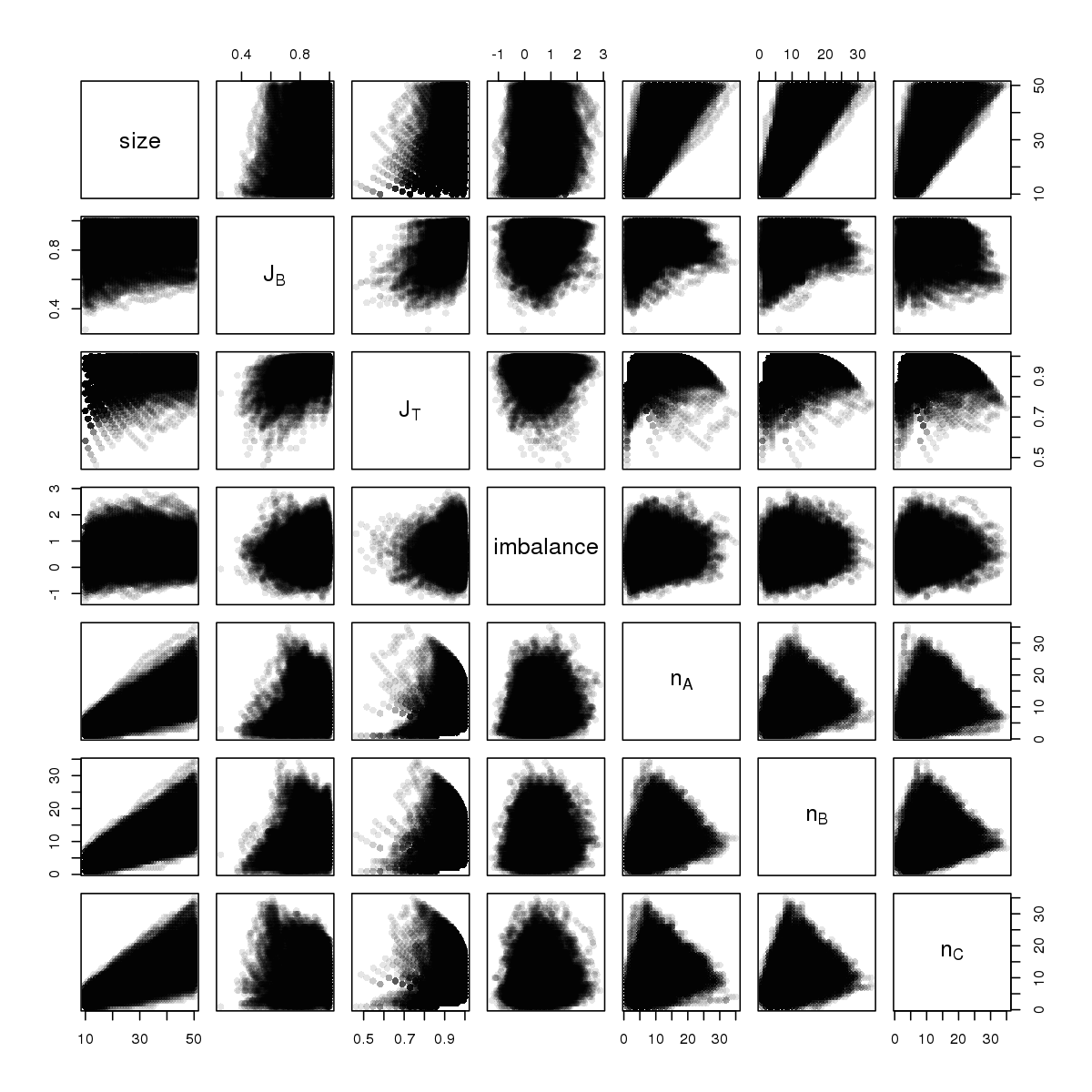} \hfill{}
\caption{
  The distribution of tree shape statistics across all subtrees included in this study.
  Tree size is number of tips.
  The $J_B$, $J_T$, and imbalance metrics are described in the text.
  $n_i$ refers to the number of tips in regime $i$, $i \in \{A,B,C\}$.
}
\label{fig:treemetrics}
\end{figure}

To get a sense of how how much variation there is in tree shape among the subtrees sampled from the large \emph{Anolis} phylogeny, we show scatterplots of tree size, our regime evenness metrics $J_B$ and $J_T$, tree imbalance \citep[the standardized Colless index,][]{Mooers1997}, and the number of tips in each regime across all 65600 subtrees used in this study (40 regime paintings $\times$ 40 subtree specifications $\times$ 40 tree sizes, Fig.~\ref{fig:treemetrics}).
These scatterplots show that there are no discernable biases in how we generated trees, in terms of the effect of e.g., size on tree imbalance or regime evenness.
There is an interesting, and not entirely unexpected, positive correlation between branch and tip evenness (correlation coefficient of 0.54).
This reflects the fact that trees with highly balanced regimes across the internal branches are more likely to have highly balanced regimes across the tips.
This correlation is at least a partial explanation for why branch evenness is a more important predictor variable than tip evenness in our GAM regressions.
We attempted to account for this correlation in our regressions, for example by using residual tip evenness (observed tip evenness minus the expectation based on the linear regression of tip evenness on branch evenness), but this did not provide a better fit to the data or even increase the statistical significance of tip evenness as an explanatory variable.

\begin{figure}
\includegraphics[width=\linewidth]{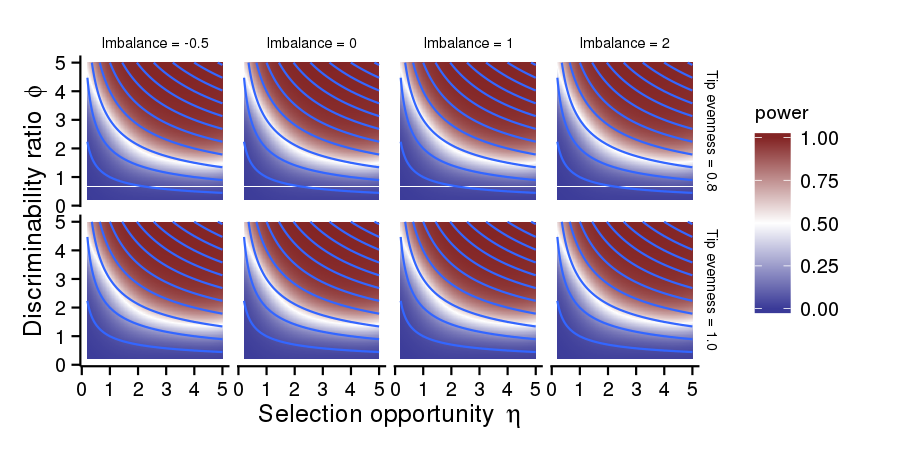} \hfill{}
\caption{
  Power as a function of selection opportunity $\eta$ and discriminability ratio $\phi$ for a range of tree imbalance and tip evenness values.
  Contours show increasing SNR.
  Predictions are based on the best-fitting GAM regression with tree size held constant at 30 and branch evenness held constant at 1.0.
}
\label{fig:eta-phi-power-contours-appendix}
\end{figure}

\begin{figure}
\includegraphics[width=\linewidth]{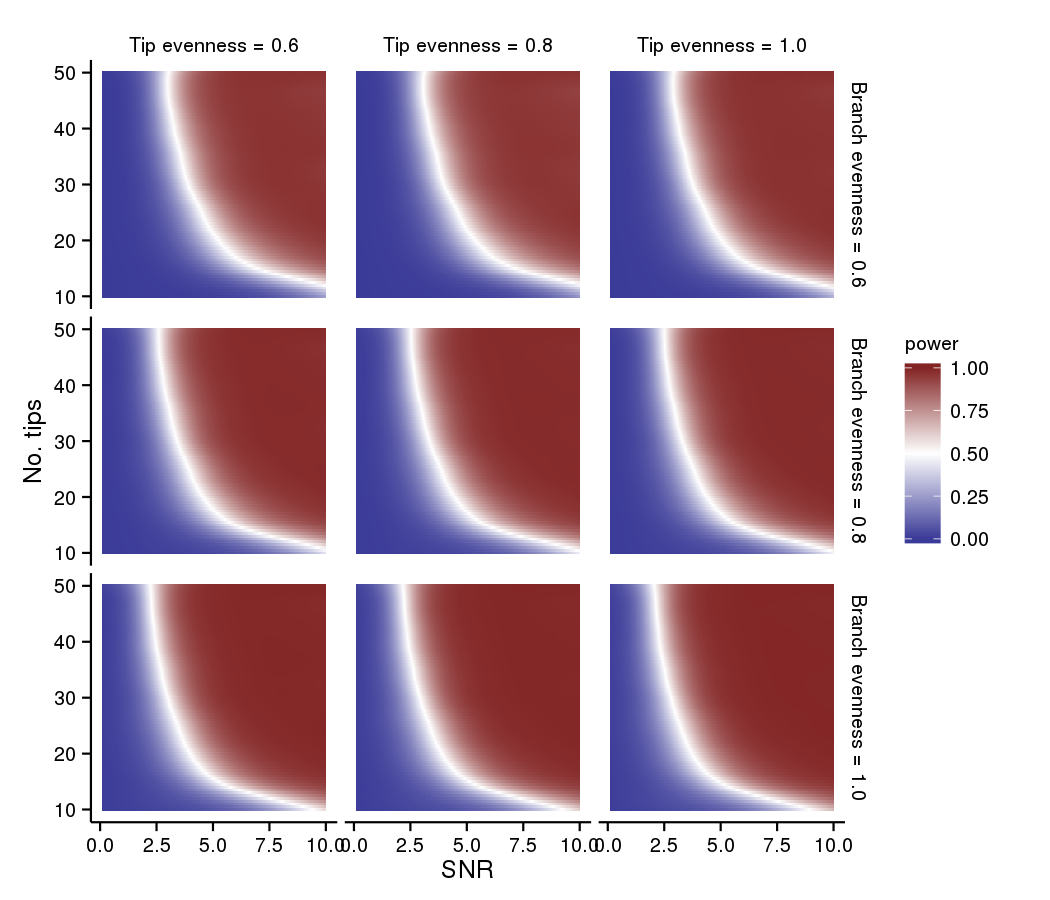} \hfill{}
\caption{
  Power as SNR and tree size are varied for a range of branch and tip evenness values.
  Power is predicted using the best-fitting GAM regression equation.
  Red corresponds to better than 50\% probability of selecting the correct model;
  blue, to less than 50\%.
}
\label{fig:snr-power-contours}
\end{figure}

Fig.~\ref{fig:eta-phi-power-contours-appendix} shows heatmaps of power across variation in tree imbalance and tip evenness values.
Tree size and branch evenness were both held constant for these figures.
The predictions are based on GAM regressions of observed power against $\eta$, $\phi$, tree size, tree imbalance, and regime evenness ($J_B$ and $J_T$).
The contours show increasing signal-to-noise ratio (SNR, $\sqrt{\eta}~\phi$).
One can see that neither tree imbalance nor tip evenness has any discernable effect on predicted power.
The significance of these two variables in the GAM regression is largely due to the large sample sizes rather than to effect size.

\begin{figure}
\includegraphics[width=\linewidth]{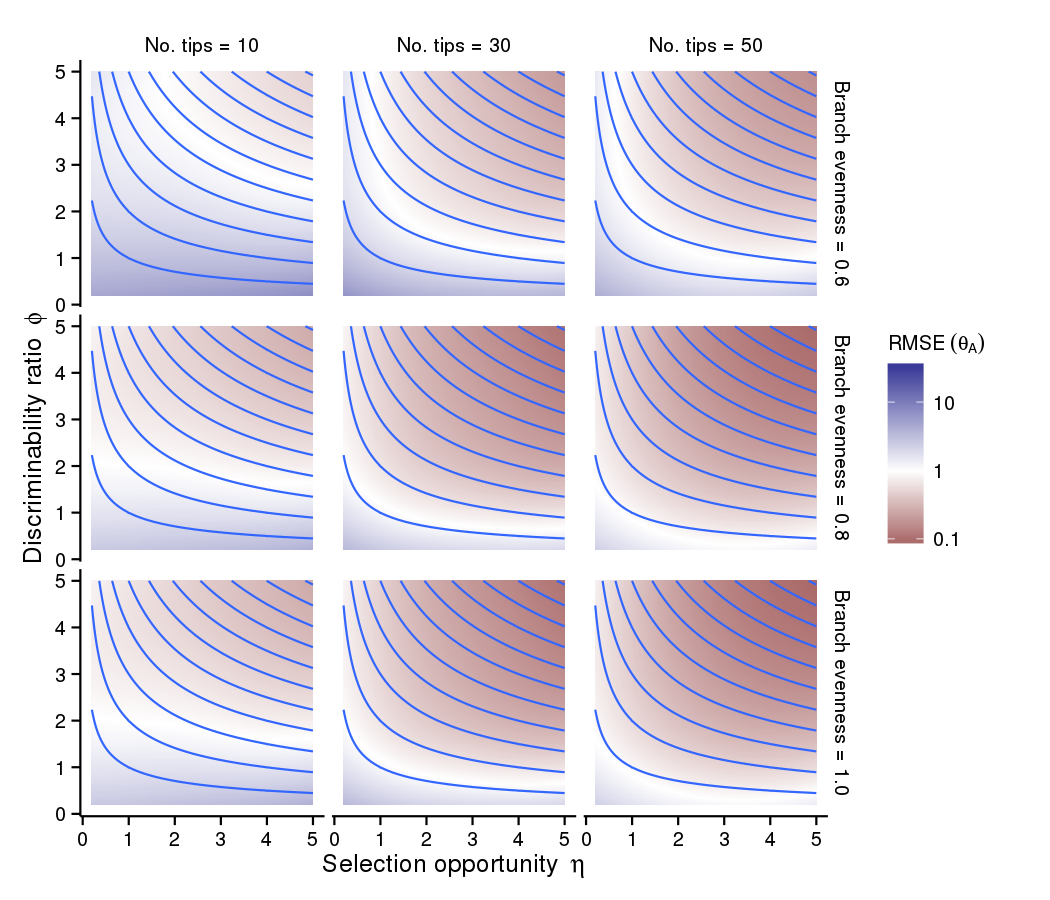} \hfill{}
\caption{The root mean square error in the estimate of the selective optimum $\theta_A$ as $\eta$ and $\phi$ are varied for a range of tree size and branch evenness values.
Contours show increasing SNR.
Predictions are based on the best-fitting GAM regression.}
\label{fig:eta-phi-thetaA-contours}
\end{figure}

\begin{figure}
\includegraphics[width=\linewidth]{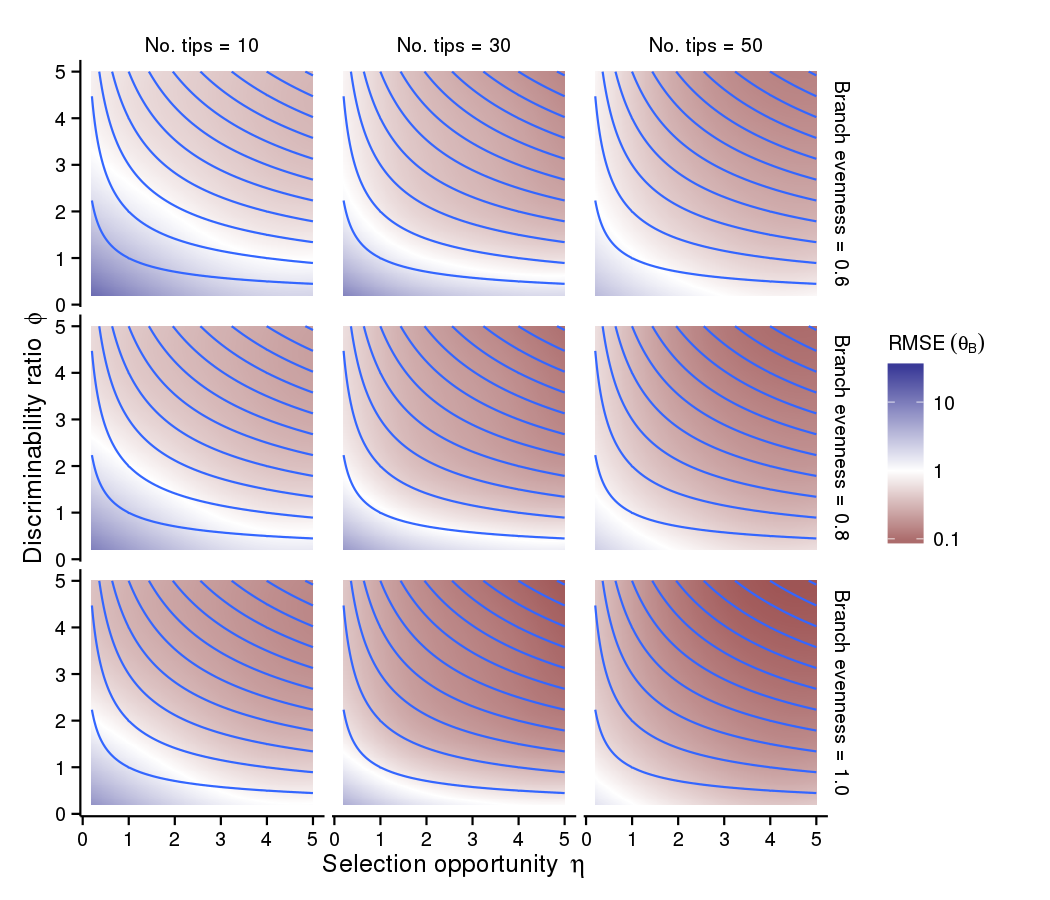} \hfill{}
\caption{The root mean square error in the estimate of the selective optimum $\theta_B$ as $\eta$ and $\phi$ are varied for a range of tree size and branch evenness values.
Contours show increasing SNR.
Predictions are based on the best-fitting GAM regression.}
\label{fig:eta-phi-thetaB-contours}
\end{figure}

\begin{figure}
\includegraphics[width=\linewidth]{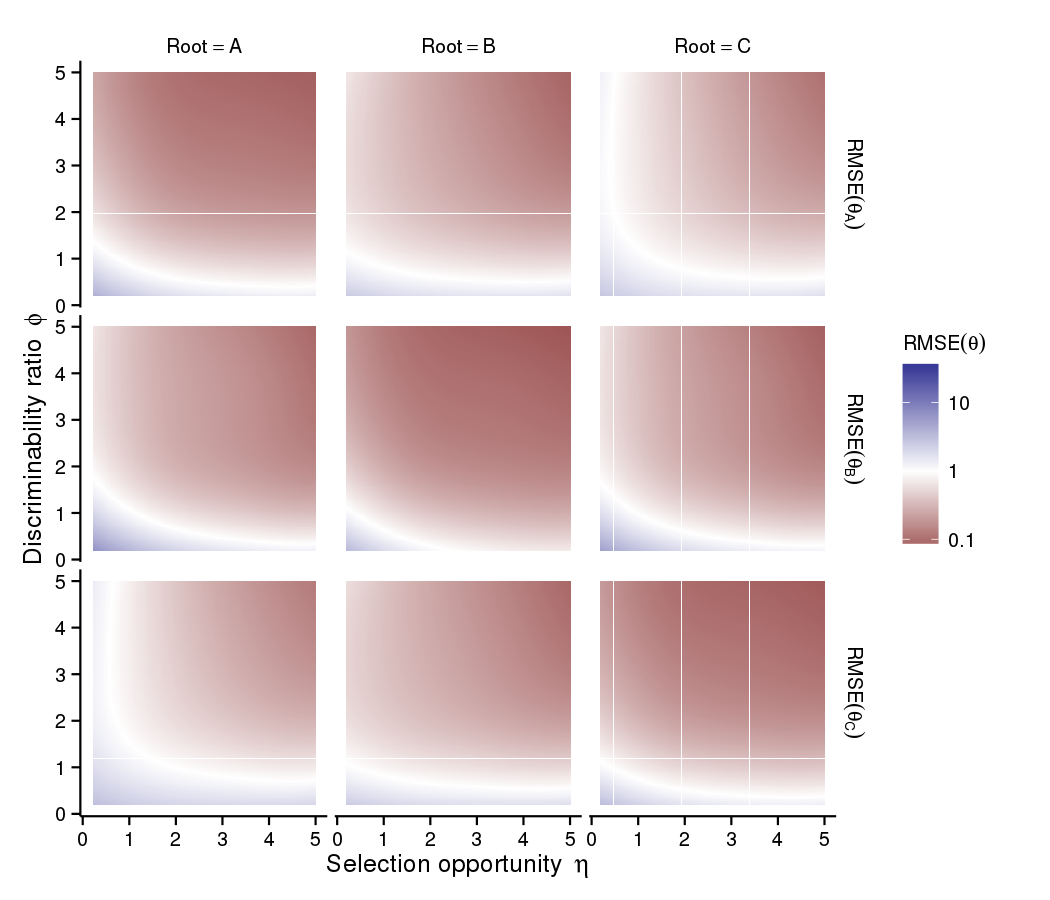} \hfill{}
\caption{
  Root mean square error in the estimate of each selective optimum across a range of $\eta$ and $\phi$ values, after accounting for the identity of the root regime.
  These heatmaps are the predictions of RMSE based on GAM regressions to subsets of the full simulation study that consider only those regime paintings where the root was in each given regime.
  For these predictions, tree size was assumed equal to 30, branch and tip evenness were assumed equal to one, and tree imbalance was assumed equal to -0.5.
  Each regime is best estimated when the root is in that regime.
}
\label{fig:theta-RMSE-by-root}
\end{figure}

Figs.~\ref{fig:eta-phi-thetaA-contours} and \ref{fig:eta-phi-thetaB-contours} show heatmaps of predicted root mean square error (RMSE) in the estimates of $\theta_A$ and $\theta_B$ as $\eta$ and $\phi$ are varied, for a range of values of tree size and branch evenness.
Tip evenness and tree imbalance were held constant.
These predictions are based on GAM regressions of observed RMSE against $\eta$, $\phi$, tree size, tree imbalance, and regime evenness ($J_B$ and $J_T$).
The contours show increasing signal-to-noise ratio (SNR, $\sqrt{\eta}~\phi$).
Comparing these plots to Fig.~\ref{fig:eta-phi-thetaC-contours} in the main text, one sees that although the response to changes in predictor variables is the same, $\theta_C$ is much better predicted than $\theta_A$ and $\theta_B$.
We discuss this result in more detail in the main paper.
It has to do with the fact that the regime present at the root is better estimated than any of the other optima;
by chance, 18 of the 40 regime paintings we used happened to have the root in regime C, compared to 12 paintings with the root in regime B and 10 with the root in regime A.
Controlling for the identity of the root regime, the response of RMSE to changes in predictor variables is nearly identical across selective optima (Fig.~\ref{fig:theta-RMSE-by-root}).

\begin{figure}
\includegraphics[width=\linewidth]{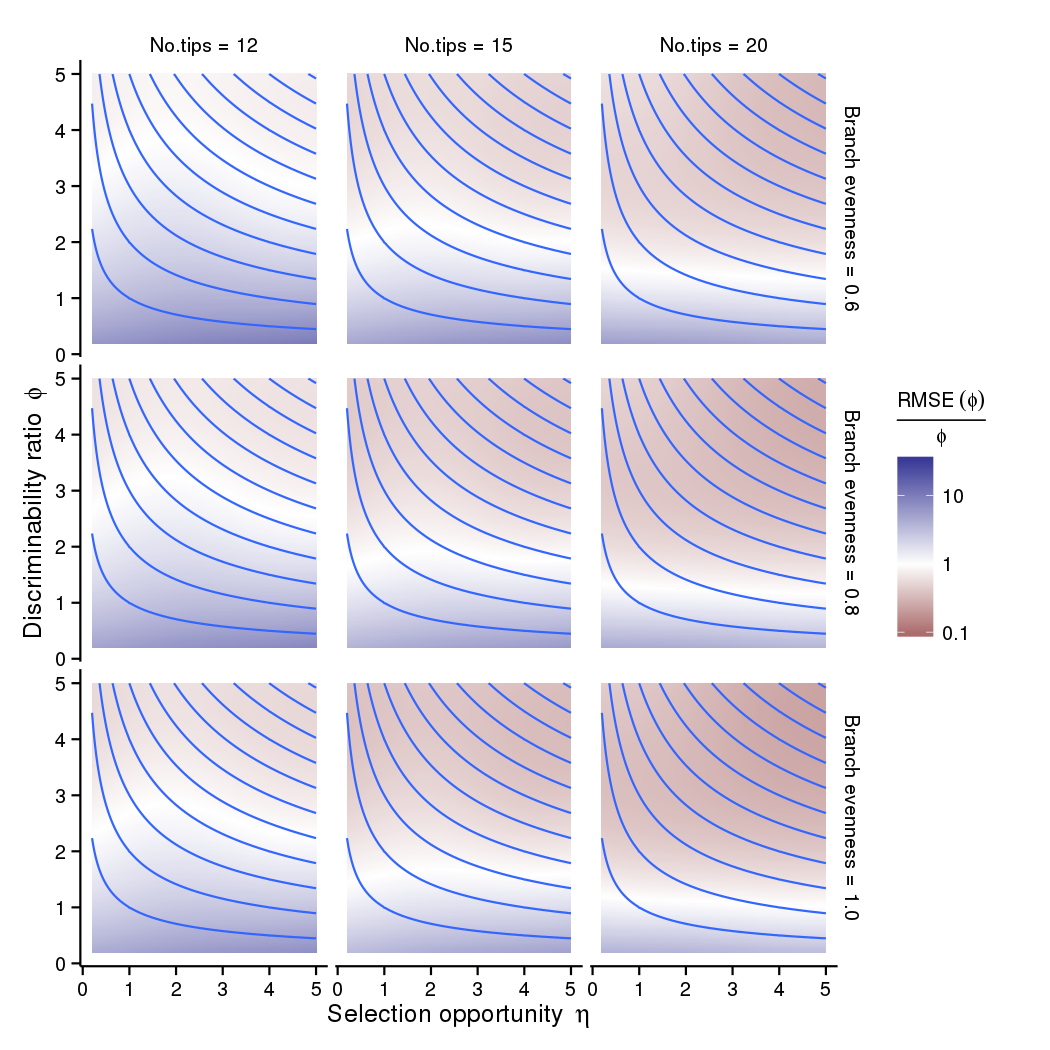} \hfill{}
\caption{The relative root mean square error in the estimate of discriminability ratio $\phi$ as $\eta$ and $\phi$ are varied for a finer range of tree size values.}
\label{fig:eta-and-phi-phi-contours-2}
\end{figure}

\begin{figure}
\includegraphics[width=\linewidth]{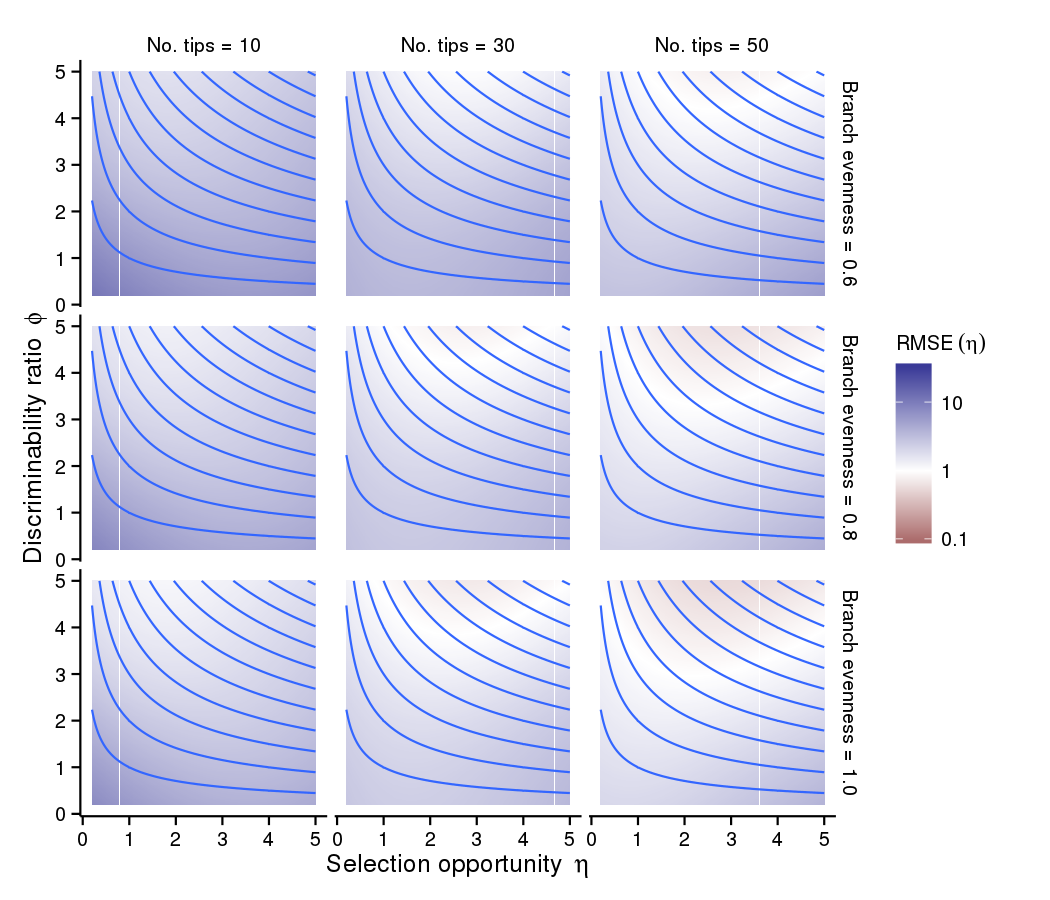} \hfill{}
\caption{The GAM-predicted absolute root mean square error in the estimate of selection opportunity $\eta$ as $\eta$ and $\phi$ are varied for a range of tree sizes and branch evenness. As in the main text, tree imbalance and tip evenness are assumed fixed at -0.5 and 1.0, respectively.}
\label{fig:eta-and-phi-abs-eta-contours}
\end{figure}

\begin{figure}
\includegraphics[width=\linewidth]{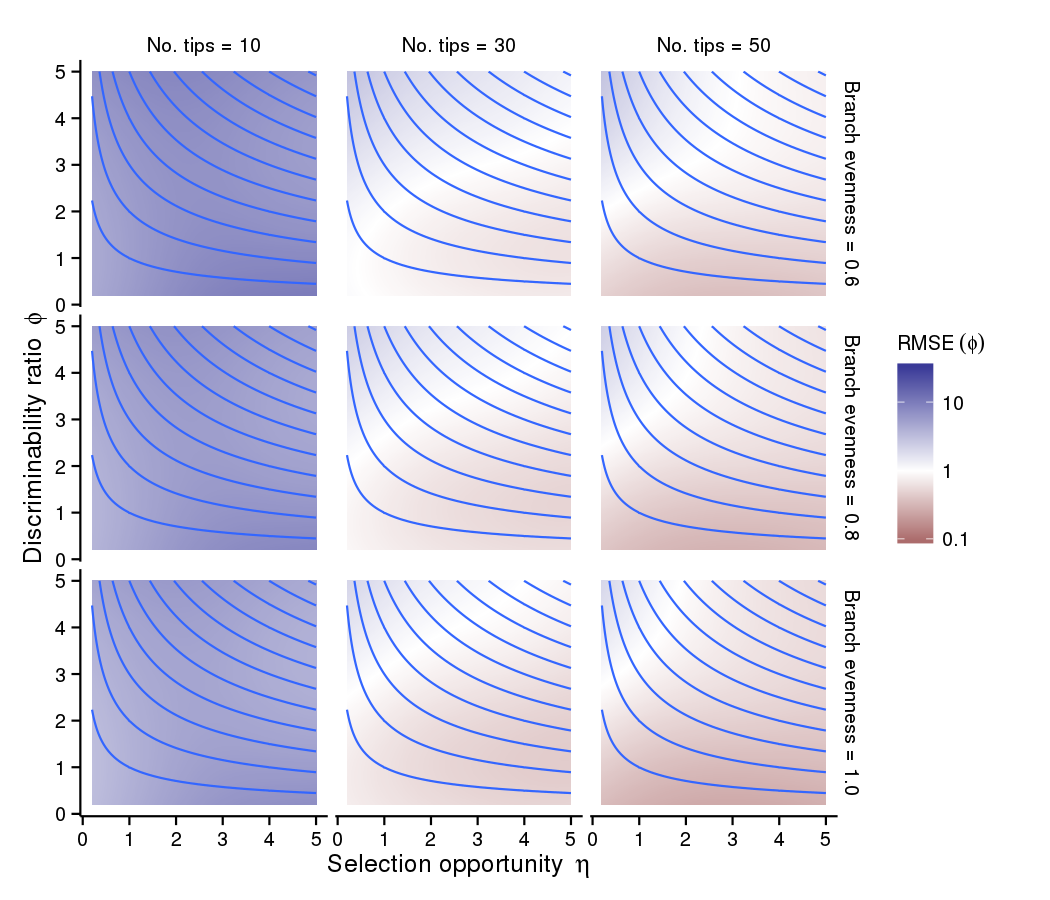} \hfill{}
\caption{The GAM-predicted absolute root mean square error in the estimate of discriminability ratio $\phi$ as $\eta$ and $\phi$ are varied for a range of tree sizes and branch evenness. As in the main text, tree imbalance and tip evenness are assumed fixed at -0.5 and 1.0, respectively.}
\label{fig:eta-and-phi-abs-phi-contours}
\end{figure}

\begin{figure}
\includegraphics[width=\linewidth]{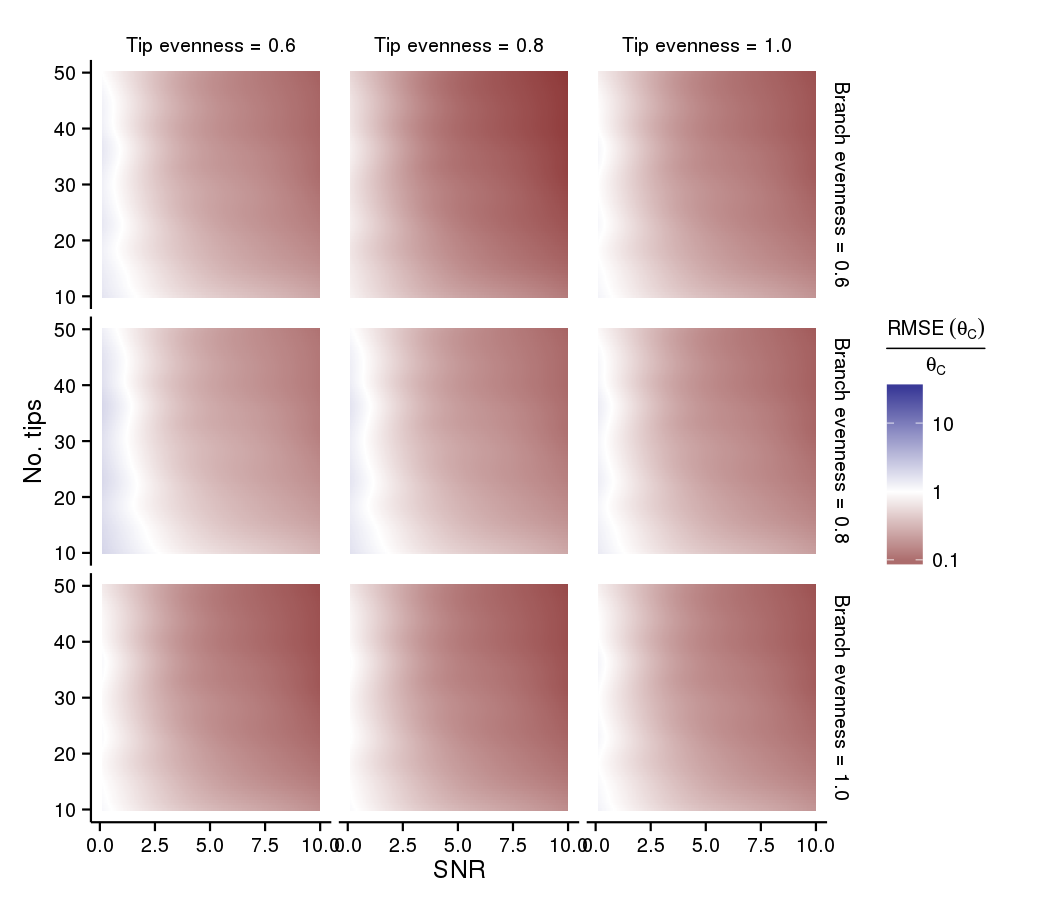} \hfill{}
\caption{
  The root mean square error in the estimate of the selective optimum $\theta_C$ as SNR and tree size are varied, for a range of tip and regime evenness values.
  Predictions are based on the best-fitting GAM regression.
}
\label{fig:snr-thetaC-contours}
\end{figure}

SNR does a reasonably good job as a predictor of the error in the estimates of the selective optima (Figs.~\ref{fig:eta-phi-thetaC-contours}, \ref{fig:eta-phi-thetaA-contours}, and \ref{fig:eta-phi-thetaB-contours}).
However, it is not a good predictor of either $\eta$ or $\phi$, as suggested by the lack of correspondence between predicted error and SNR in Figs.~\ref{fig:eta-phi-phi-contours} and \ref{fig:eta-phi-eta-contours}.
These conclusions are supported by comparing the $R^2$ values of the GAM regressions using SNR instead of $\eta$ and $\phi$ separately ($\eta$ regression $R^2$ drops from 0.66 to 0.20, $\phi$ regression $R^2$ drops from 0.69 to 0.48, and $\theta_C$ regression $R^2$ drops from 0.45 to 0.43).
Fig.~\ref{fig:snr-thetaC-contours} shows RMSE in the estimate of $\theta_C$ as SNR and tree size varies, for a range of tip and regime evenness values.

\clearpage
\newpage
\section{Generalized additive model regressions}
\label{sec:gams}
\setcounter{table}{0}
\renewcommand{\thetable}{B\arabic{table}}

We fit generalized additive models (GAMs) to capture the dependence of observed power and mean square error on the simulation study parameters ($\eta$, $\phi$, tree size, tree imbalance, and regime evenness).
In the GAM framework, the location of the response variable is assumed to be a smooth function of the predictor variables.
GAMs are appropriate here because we had no expectation as to the shape of the dependence of the response variable on any of the predictors.
The GAMs accomodate this need for flexibility using thin plate regression splines to describe the relationship between the response and the predictors \citep{Wood2003,Wood2006}.
Thin plate regression splines are simply splines with two or more basis dimensions, where the ``wiggliness'' of the spline is penalized to avoid overfitting.
A nice physical interpretation of thin plate regression splines is to visualize the relationship between the response and two predictor variables as a rugged landscape where the height of the landscape gives the value of the response variable at combinations of predictor variables.
The thin plate regression spline can be visualized as a thin sheet of metal that is being bent to fit the shape of this landscape;
the rigidity of the metal is analogous to the penalty applied to the spline to avoid overfitting.
Fits were accomplished using the \texttt{gam} function provided in the R package \texttt{mgcv} \citep{Wood2006}.
This function allows different predictors to be added, removed, or combined and the resulting fits to be compared via analysis of deviance, as with generalized linear models.

In the tables below, we display analysis of deviance tables comparing several GAMs for each response variable (power, relative RMSE in the estimates of $\eta$ and $\phi$, RMSE in the estimates of the selective optima).
There are a number of conclusions we can draw.
First, the best-fitting model was the same for all response variables.
This model predicts the response as an additive combination of smooths:
a two-dimensional smooth of branch and tip evenness;
a one-dimensional smooth of tree imbalance;
and a three-dimensional smooth of the simulation parameters ($\eta, \phi$, and tree size).
Second, the largest improvement in model fit comes from adding branch evenness as an additive fixed effect.
There is a significant improvement from adding fixed effects of tree imbalance and tip evenness, but these only slightly improve model $R^2$.
Allowing the response to be a smooth function of the tree shape and regime evenness metrics improves the fit, especially for the estimates of the selective optima.

We also compare the $R^2$ for the best-fitting models using selection opportunity $\eta$ and discriminability ratio $\phi$ as separate predictors to the best-fitting models using the composite parameter of signal-to-noise ratio (SNR: $\sqrt{\eta}~\phi$) as a predictor (Table~\ref{tab:eta-phi-vs-snr}).
It is clear from this comparison that models using SNR as a predictor perform nearly as well as models with $\eta$ and $\phi$ separately for power, but that the ability to predict estimation accuracy is substantially worse.

\begin{table}[ht]
  \small
  \caption{
    Comparison of GAM regression fits of observed power $p$.
    Power was logit-transformed prior to analysis.
    This transformation allows the response to vary between $(-\infty,\infty)$ instead of being bounded between 0 and 1, making it easier to fit the GAM to the data.
  }
\begin{tabular}{lrrrrr}
  \hline
 & Resid.~Df & Resid.~Dev & Df & Deviance & $R^2$\\
  \hline
  $\log\left(\frac{p}{1-p}\right)$\textasciitilde s(be,te)+s(imb)+s($\eta,\phi$,size) & 351854 & 206658.00 &  &  & 0.93866 \\
  $\log\left(\frac{p}{1-p}\right)$\textasciitilde s(be)+s(te)+s(imb)+s($\eta,\phi$,size) & 351865 & 206831.00 & -12 & -173.78 & 0.93861 \\
  $\log\left(\frac{p}{1-p}\right)$\textasciitilde be+te+imb+s($\eta,\phi$,size) & 351887 & 207431.00 & -22 & -599.84 & 0.93844 \\
  $\log\left(\frac{p}{1-p}\right)$\textasciitilde be+imb+s($\eta,\phi$,size) & 351888 & 207494.00 & -1 & -63.08 & 0.93832 \\
  $\log\left(\frac{p}{1-p}\right)$\textasciitilde be+s($\eta,\phi$,size) & 351889 & 207836.00 & -1 & -341.79 & 0.93832\\
  $\log\left(\frac{p}{1-p}\right)$\textasciitilde s($\eta,\phi$,size) & 351890 & 256837.00 & -1 & -49000.90 & 0.92378 \\
   \hline
\end{tabular}
\end{table}

\begin{table}[ht]
  \small
  \caption{Comparison of GAM regression fits of observed relative root mean square error in the estimate of selection opportunity $\eta$.
Relative RMSE was log-transformed to deal with the occasional large value---the maximum observed RMSE was greater than $10^3$.} \begin{tabular}{lrrrrr}
  \hline
 & Resid.~Df & Resid.~Dev & Df & Deviance & $R^2$ \\
  \hline
  $\log_{10}\left(\frac{RMSE(\eta)}{\eta}\right)$\textasciitilde s(be,te)+s(imb)+s($\eta,\phi$,size) & 271086 & 13608 &  & &0.65608 \\
  $\log_{10}\left(\frac{RMSE(\eta)}{\eta}\right)$\textasciitilde s(be)+s(te)+s(imb)+s($\eta,\phi$,size) & 271097 & 13619 & -10.43 & -11.89 & 0.65579\\
  $\log_{10}\left(\frac{RMSE(\eta)}{\eta}\right)$\textasciitilde be+te+imb+s($\eta,\phi$,size) & 271118 & 13650 & -22 & -21.84 & 0.65505 \\
  $\log_{10}\left(\frac{RMSE(\eta)}{\eta}\right)$\textasciitilde be+imb+s($\eta,\phi$,size) & 271119 & 13650.10 & -1 & -0.15 & 0.65504 \\
  $\log_{10}\left(\frac{RMSE(\eta)}{\eta}\right)$\textasciitilde be+s($\eta,\phi$,size) & 271120 & 13669.90 & -1 & -19.73 & 0.65455 \\
  $\log_{10}\left(\frac{RMSE(\eta)}{\eta}\right)$\textasciitilde s($\eta,\phi$,size) & 271121 & 13792.50 & -1 & -122.63 & 0.65145 \\
   \hline
\end{tabular}
\end{table}

\begin{table}[ht]
  \small
  \caption{Comparison of GAM regression fits of observed relative root mean square error in the estimate of discriminability ratio $\phi$.
Relative RMSE was log-transformed to deal with the occasional large value---the maximum observed RMSE was greater than 80.} \begin{tabular}{lrrrrr}
  \hline
 & Resid.~Df & Resid.~Dev & Df & Deviance & $R^2$ \\
  \hline
  $\log_{10}\left(\frac{RMSE(\phi)}{\phi}\right)$\textasciitilde s(be,te)+s(imb)+s($\eta,\phi$,size) & 271087 & 5417.5 &  &  & 0.68922 \\
  $\log_{10}\left(\frac{RMSE(\phi)}{\phi}\right)$\textasciitilde s(be)+s(te)+s(imb)+s($\eta,\phi$,size) & 271097 & 5419.3 & -10.76 & -1.87 &  0.68913 \\
  $\log_{10}\left(\frac{RMSE(\phi)}{\phi}\right)$\textasciitilde be+te+imb+s($\eta,\phi$,size) & 271118 & 5438.3 & -20.56 & -18.95 & 0.68806 \\
  $\log_{10}\left(\frac{RMSE(\phi)}{\phi}\right)$\textasciitilde be+imb+s($\eta,\phi$,size) & 271119 & 5438.4 & -1 & -0.067 & 0.68806 \\
  $\log_{10}\left(\frac{RMSE(\phi)}{\phi}\right)$\textasciitilde be+s($\eta,\phi$,size) & 271120 & 5438.5 & -1 & -0.17 & 0.68805 \\
   $\log_{10}\left(\frac{RMSE(\phi)}{\phi}\right)$\textasciitilde s($\eta,\phi$,size) & 271121 & 5662.3 & -1 & -223.76 & 0.67522 \\
   \hline
\end{tabular}
\end{table}

\begin{table}[ht]
  \small
  \caption{Comparison of GAM regression fits of observed relative root mean square error in the estimate of the selective optimum $\theta_A$.
Relative RMSE was log-transformed to deal with the occasional large value---the maximum observed RMSE was greater than 35.} \begin{tabular}{lrrrrr}
  \hline
 & Resid.~Df & Resid.~Dev & Df & Deviance & $R^2$ \\
  \hline
  $\log_{10}\left(RMSE(\theta_A)\right)$\textasciitilde s(be, te) + s(imb) + s($\eta,\phi$,size) & 271085 & 20900 &  & & 0.47319 \\
  $\log_{10}\left(RMSE(\theta_A)\right)$\textasciitilde s(be) + s(te) + s(imb) + s($\eta,\phi$,size) & 271096 & 20985 & -10.84 & -84.55 & 0.47108 \\
  $\log_{10}\left(RMSE(\theta_A)\right)$\textasciitilde be + te + imb + s($\eta,\phi$,size) & 271119 & 21820 & -23.58 & -835.88 & 0.45006 \\
  $\log_{10}\left(RMSE(\theta_A)\right)$\textasciitilde be + imb + s($\eta,\phi$,size) & 271120 & 21822 & -1 & -1.69 & 0.45001 \\
  $\log_{10}\left(RMSE(\theta_A)\right)$\textasciitilde be + s($\eta,\phi$,size) & 271121 & 21822 & -1 & -0.16 & 0.45001 \\
  $\log_{10}\left(RMSE(\theta_A)\right)$\textasciitilde s($\eta,\phi$,size) & 271122 & 22174 & -1 & -351.74 & 0.44115 \\
   \hline
\end{tabular}
\end{table}

\begin{table}[ht]
  \small
  \caption{Comparison of GAM regression fits of observed relative root mean square error in the estimate of the selective optimum $\theta_B$.
Relative RMSE was log-transformed to deal with the occasional large value---the maximum observed RMSE was greater than 40.}
\begin{tabular}{lrrrrr}
  \hline
 & Resid.~Df & Resid.~Dev & Df & Deviance & $R^2$\\
  \hline
$\log_{10}\left(RMSE(\theta_B)\right)$\textasciitilde s(be, te) + s(imb) + s($\eta,\phi$,size) & 271085 & 13104 &  &  & 0.53186 \\
  $\log_{10}\left(RMSE(\theta_B)\right)$\textasciitilde s(be) + s(te) + s(imb) + s($\eta,\phi$,size) & 271095 & 13212 & -10.35 & -107.48 & 0.52804 \\
  $\log_{10}\left(RMSE(\theta_B)\right)$\textasciitilde be + te + imb + s($\eta,\phi$,size) & 271118 & 13394 & -23.52 & -182.43 & 0.52157 \\
  $\log_{10}\left(RMSE(\theta_B)\right)$\textasciitilde be + imb + s($\eta,\phi$,size) & 271119 & 13460 & -1 & -65.92 & 0.51921 \\
  $\log_{10}\left(RMSE(\theta_B)\right)$\textasciitilde be + s($\eta,\phi$,size) & 271120 & 13460 & -1 & -0.76 & 0.51919 \\
  $\log_{10}\left(RMSE(\theta_B)\right)$\textasciitilde s($\eta,\phi$,size) & 271121 & 13607 & -1 & -146.68 & 0.51395 \\
   \hline
\end{tabular}
\end{table}

\begin{table}[ht]
  \small
  \caption{Comparison of GAM regression fits of observed relative root mean square error in the estimate of the selective optimum $\theta_C$.
Relative RMSE was log-transformed to deal with the occasional large value---the maximum observed RMSE was greater than 34.}
\begin{tabular}{lrrrrr}
  \hline
 & Resid.~Df & Resid.~Dev & Df & Deviance & $R^2$\\
  \hline
$\log_{10}\left(RMSE(\theta_B)\right)$\textasciitilde s(be, te) + s(imb) + s($\eta,\phi$,size) & 271093 & 20598 &  & & 0.45232 \\
  $\log_{10}\left(RMSE(\theta_B)\right)$\textasciitilde s(be) + s(te) + s(imb) + s($\eta,\phi$,size) & 271096 & 20685 & -3.17 & -86.81 & 0.45002 \\
   $\log_{10}\left(RMSE(\theta_B)\right)$\textasciitilde be + te + imb + s($\eta,\phi$,size) & 271119 & 21398 & -22.74 & -712.93 & 0.43111 \\
  $\log_{10}\left(RMSE(\theta_B)\right)$\textasciitilde be + imb + s($\eta,\phi$,size) & 271120 & 21406 & -1 & -7.69 & 0.43091 \\
  $\log_{10}\left(RMSE(\theta_B)\right)$\textasciitilde be + s($\eta,\phi$,size) & 271121 & 21415 & -1 & -8.97 & 0.43067 \\
  $\log_{10}\left(RMSE(\theta_B)\right)$\textasciitilde s($\eta,\phi$,size) & 271122 & 21442 & -1 & -26.96 & 0.42995 \\
   \hline
\end{tabular}
\end{table}

\begin{table}[ht]
\caption{Comparison of $R^2$ values for GAM models using $\eta$ and $\phi$ as predictors to GAM models using the composite parameter SNR as predictor.
\label{tab:eta-phi-vs-snr}}
\begin{tabular}{lrr}
\hline
Response & $\eta$-$\phi$ model $R^2$ & SNR model $R^2$ \\
\hline
Power & 0.93866 & 0.93669 \\
$\log_{10}\left(\frac{RMSE(\eta)}{\eta}\right)$ & 0.65580 & 0.20236 \\
$\log_{10}\left(\frac{RMSE(\phi)}{\phi}\right)$ & 0.68919 & 0.48645 \\
$\log_{10}\left(RMSE(\theta_A)\right)$ & 0.47277 & 0.43562 \\
$\log_{10}\left(RMSE(\theta_B)\right)$ & 0.53160 & 0.47181 \\
$\log_{10}\left(RMSE(\theta_C)\right)$ & 0.45232 & 0.43272 \\
\hline
\end{tabular}
\end{table}

\end{document}